\documentclass[twocolumn,floatfix,prl,aps,showpacs]{revtex4-1}
\usepackage{graphicx,amsmath,amssymb,color}
\usepackage[utf8]{inputenc}
\usepackage{nicefrac}
\usepackage{multirow, makecell, ragged2e}
\usepackage[titletoc,title]{appendix}

\newcommand{\be}{\begin{equation}}
\newcommand{\ee}{\end{equation}}

\newcommand{\ba}{\begin{eqnarray}}
\newcommand{\ea}{\end{eqnarray}}

\begin{document}

\title{
  Prediction of a non-Abelian fractional quantum Hall state with $f$-wave pairing of composite fermions in wide quantum wells
}
\author{W. N. Faugno$^1$,  Ajit C. Balram$^2$, Maissam Barkeshli$^3$, and J. K. Jain$^1$}

\affiliation{$^1$Department of Physics, 104 Davey Lab, Pennsylvania State University, University Park, Pennsylvania 16802, USA}
\affiliation{$^2$Niels Bohr International Academy and the Center for Quantum Devices, Niels Bohr Institute, University of Copenhagen, 2100 Copenhagen, Denmark}
\affiliation{$^3$Condensed Matter Theory Center and Joint Quantum Institute, Department of Physics, University of Maryland, College Park, Maryland 20472 USA}

\date{\today}

\begin{abstract} 
We theoretically investigate the nature of the state at quarter filled lowest Landau level and predict that, as the quantum well width is increased, a transition occurs from the composite fermion Fermi sea into a novel non-Abelian fractional quantum Hall state that is topologically equivalent to $f$-wave pairing of composite fermions. This state is topologically distinct from the familiar $p$-wave paired Pfaffian state. We compare our calculated phase diagram with experiments and make predictions for many  observable quantities. 
\end{abstract}
\maketitle

The Moore-Read (MR) Pfaffian model~\cite{Moore91} for the even-denominator fractional quantum Hall effect (FQHE) at filling factor $\nu=5/2$~\cite{Willett87} predicts Majorana excitations that are neither fermionic nor bosonic but obey non-Abelian braid statistics~\cite{Read00}. This follows most directly from the understanding that the MR wave function represents a topological chiral $p$-wave ``superconductor" of composite fermions~\cite{Read00}, which themselves are emergent particles formed from the binding of electrons and quantized vortices~\cite{Jain89,Jain07}. Quasiparticle tunneling~\cite{Radu08}, quasiparticle interference~\cite{Willett09,Willett10}, and thermal Hall~\cite{Banerjee17,Banerjee17b} experiments have sought to measure the Majorana excitations, but the observations are not fully consistent with the predictions arising from either the Pfaffian~\cite{Moore91} or its hole conjugate called the anti-Pfaffian~\cite{Levin07,Lee07}. Realization of other non-Abelian states will therefore not only be fundamentally interesting in its own right, but can help provide an unambiguous demonstration of non-Abelian anyons. We predict in this Letter that the FQHE state observed at $\nu=1/4$ in wide quantum wells (WQWs)~\cite{Luhman08,Shabani09a,Shabani09b,Shabani13} provides a realization of a new type of non-Abelian state~\cite{Jain89b,Jain90} that is  topologically distinct from the (anti-)Pfaffian state. We make detailed predictions for several topological properties of this state that are measurable by currently available experimental techniques.

The $\nu=1/4$ state of our interest belongs to a large class of states appearing within the parton theory of the FQHE~\cite{Jain89b,Jain90}. Here one divides each electron into $k$ fictitious particles called partons, places each species of parton into an integer quantum Hall effect (IQHE) state with filling $n_\lambda$, and then glues the partons back together to recover the physical electrons.  This leads to candidate ``$n_1 n_2\cdots n_k$" FQHE states~\cite{Jain89b,Jain90}:
\be
\Psi^{n_1 n_2\cdots n_k} = {\cal P}_{\rm LLL}\prod_{\lambda=1}^k \Phi_{n_\lambda}(\{z_{i}\}).
\label{eq:parton}
\ee
Here $\Phi_n$ is the wave function of the IQHE state with $n$ filled Landau levels (LLs), $\{z_{i}=x_i-y_i\}$ are electron coordinates, and ${\cal P}_{\rm LLL}$ represents projection into the lowest LL (LLL). Negative values of $n$ are denoted as $\bar{n}$, with $\Phi_{-n}=\Phi_{\bar{n}}\equiv [\Phi_n]^*$ being the wave function of $|n|$ filled LL state in a negative magnetic field. To ensure equal area for each parton species, the charge is given by $e_\lambda=\nu/n_\lambda$ in units of the electron charge, with $\sum_{\lambda=1}^k e_\lambda=1$.  The candidate wave function $\Psi^{n_1 n_2\cdots n_k}$ represents an incompressible state at filling factor $\nu = \left[\sum_{\lambda=1}^k n_\lambda^{-1}\right]^{-1}$.  Remarkably, even though the partons themselves are unphysical, they leave their footprints in the physical world; for example, an excited parton in the factor $\Phi_{n_\lambda}$ produces a charge $e_\lambda$ excitation in the physical state. A field theoretical description of these states was constructed by Wen and collaborators~\cite{Blok90,Blok90b,Wen91,Wen92b}.

The familiar wave functions of the composite-fermion (CF) theory $\Psi_{\nu=n/(2pn+1)} = {\cal P}_{\rm LLL}\Phi_n \Phi_1^{2p}$ and $ \Psi_{\nu=n/(2pn-1)} = {\cal P}_{\rm LLL}\Phi_{\bar{n}} \Phi_1^{2p}$ are obtained as $n11\cdots$ and $\bar{n}11\cdots$ states. The parton theory contains states beyond the CF theory. Wen showed~\cite{Wen91} that the Jain parton states of the form $\Psi_{\nu=n/k}^{nn\cdots}=[\Phi_n]^k$ with $n\geq 2$ and $k\geq 2$ are non-Abelian. For these states, because all $k$ partons are identical, the theory must be invariant under an $SU(k)$ rotation within the internal parton space. Imposing this constraint through a non-Abelian gauge field and integrating out the partons leads to an $SU(k)_n$ Chern-Simons theory, which implies that the underlying states hosts non-Abelian quasiparticles.  Wen showed~\cite{Wen91} that the $[\Phi_n]^k$ state has chiral central charge $c=n(kn+1)/(k+n)$. In particular, the bosonic 22 state at $\nu=1$ has $c=5/2$. Other states that contain factors of $[\Phi_n]^k$ also support non-Abelian quasiparticles for the same reason. The electron states 221 at $\nu=1/2$ and 22111 at $\nu=1/4$, described by $U(1)\times SU(2)_2$ Chern-Simons theory, also are non-Abelian with $c=5/2$. 

All wave functions in Eq.~(\ref{eq:parton}) are in principle valid candidates for FQHE, but the important question is which ones occur for realistic interactions. Extensive work has shown that the LLL primarily stabilizes composite fermions. For states beyond the CF theory, one must therefore look to higher LLs, to monolayer or bilayer graphene, or to systems in WQWs, all of which have different Coulomb matrix elements than purely two-dimensional electrons in the LLL.  The simplest non-Abelian parton state, namely the 221 state at $\nu=1/2$~\cite{Jain89b,Jain90,Wen91,Wu16,Bandyopadhyay18}, is not a satisfactory candidate for the $\nu=1/2$ FQHE in the second LL, i.e., the 5/2 FQHE, because exact diagonalization does not produce an incompressible state at the corresponding ``shift"~\cite{Wojs09}. Recently, Balram, Barkeshli and Rudner~\cite{Balram18} have shown the surprising result that the seemingly more complicated $\bar{2}\bar{2}111$ state provides a rather good representation of the Coulomb ground state at $\nu=5/2$, although the $\bar{2}\bar{2}111$ state happens to lie in the same universality class as the anti-Pfaffian state. There are indications that the 221 state may be relevant to 1/2 FQHE in bilayer graphene for appropriate parameters~\cite{Wu16} and to the $n=3$ LL of monolayer graphene~\cite{Kim18}. 

\begin{figure*}[t]
		\includegraphics[width = 2.3in]{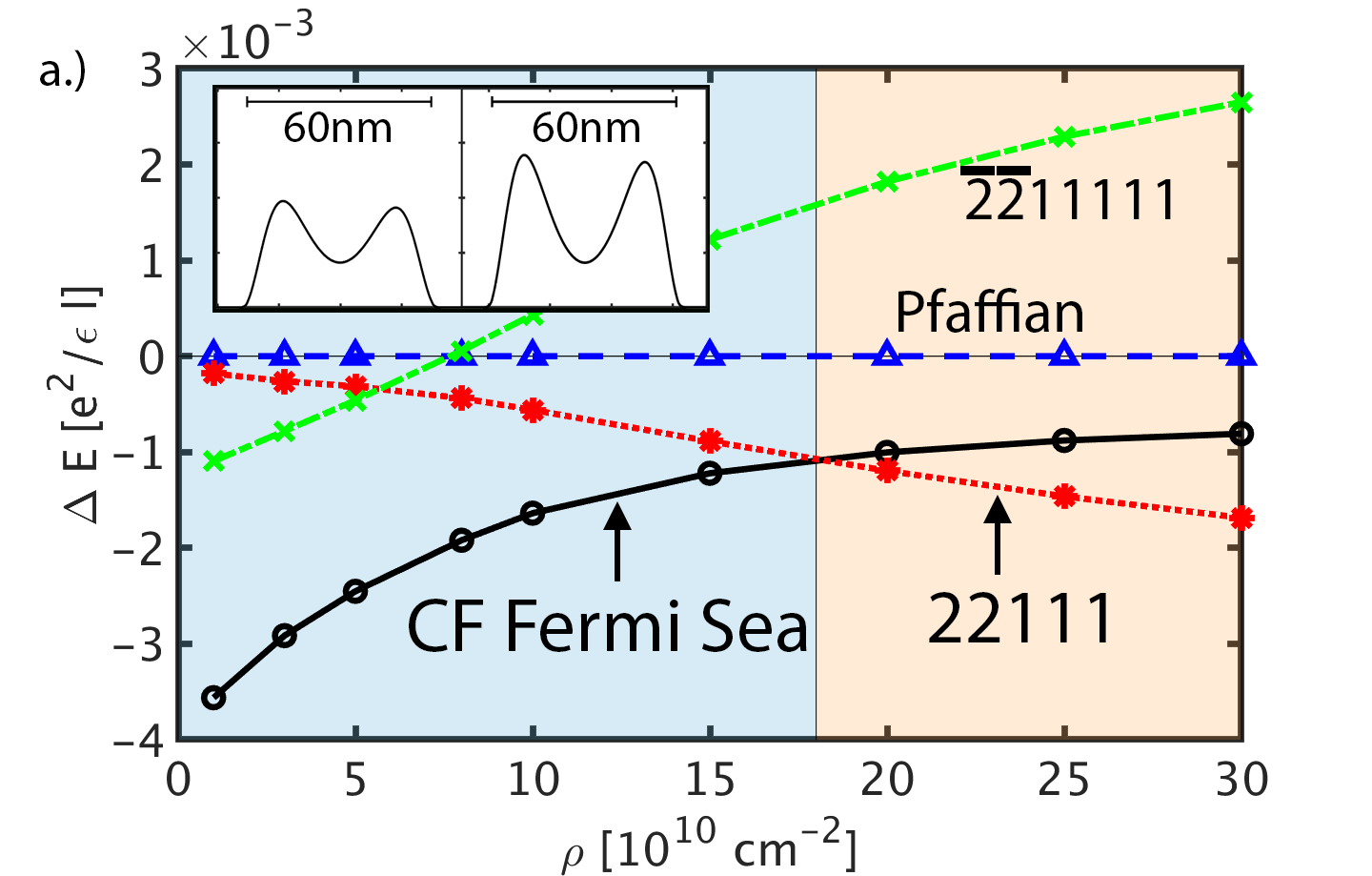}
		\includegraphics[width = 2.3in]{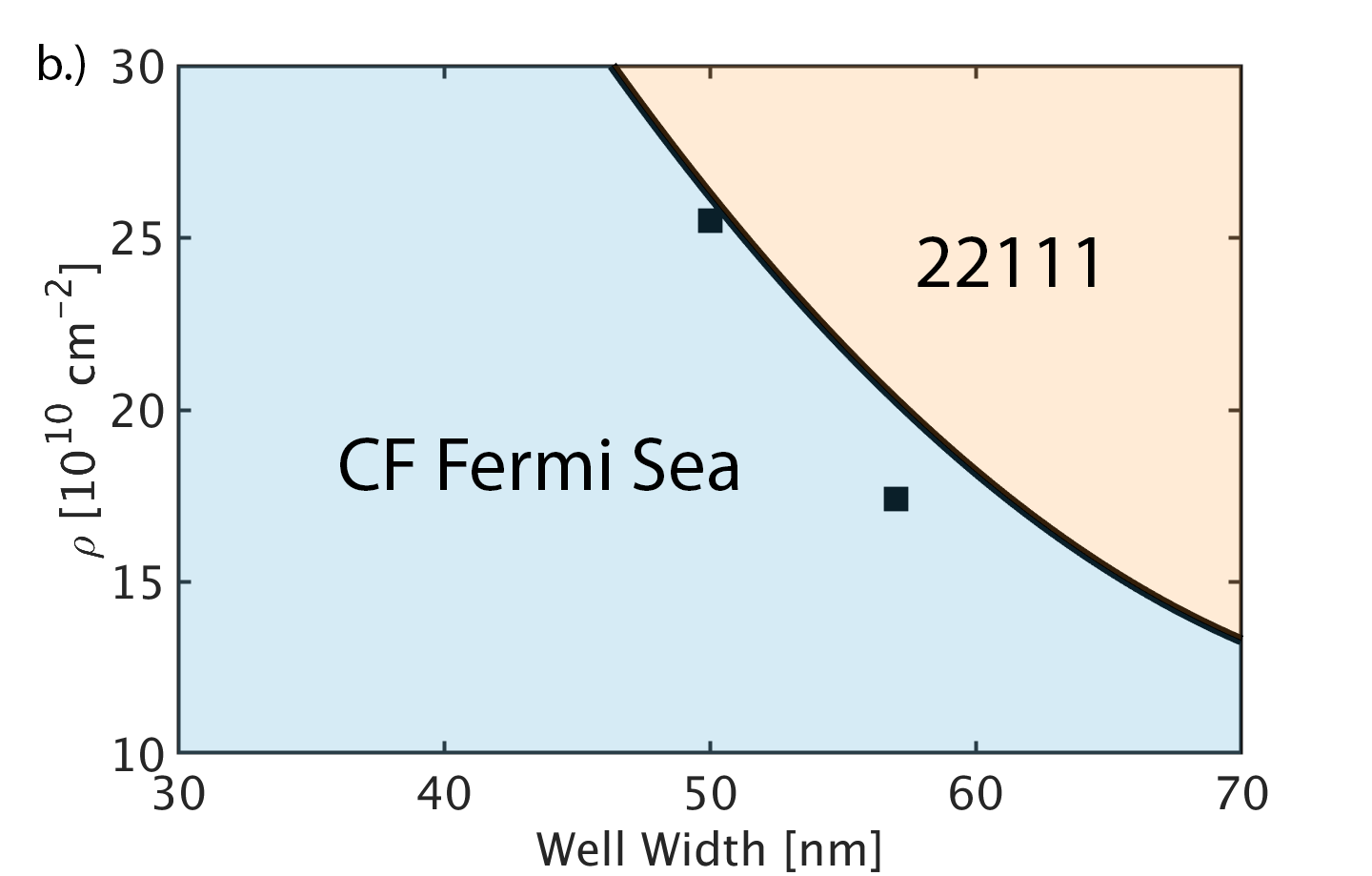}
		\includegraphics[width = 2.3in]{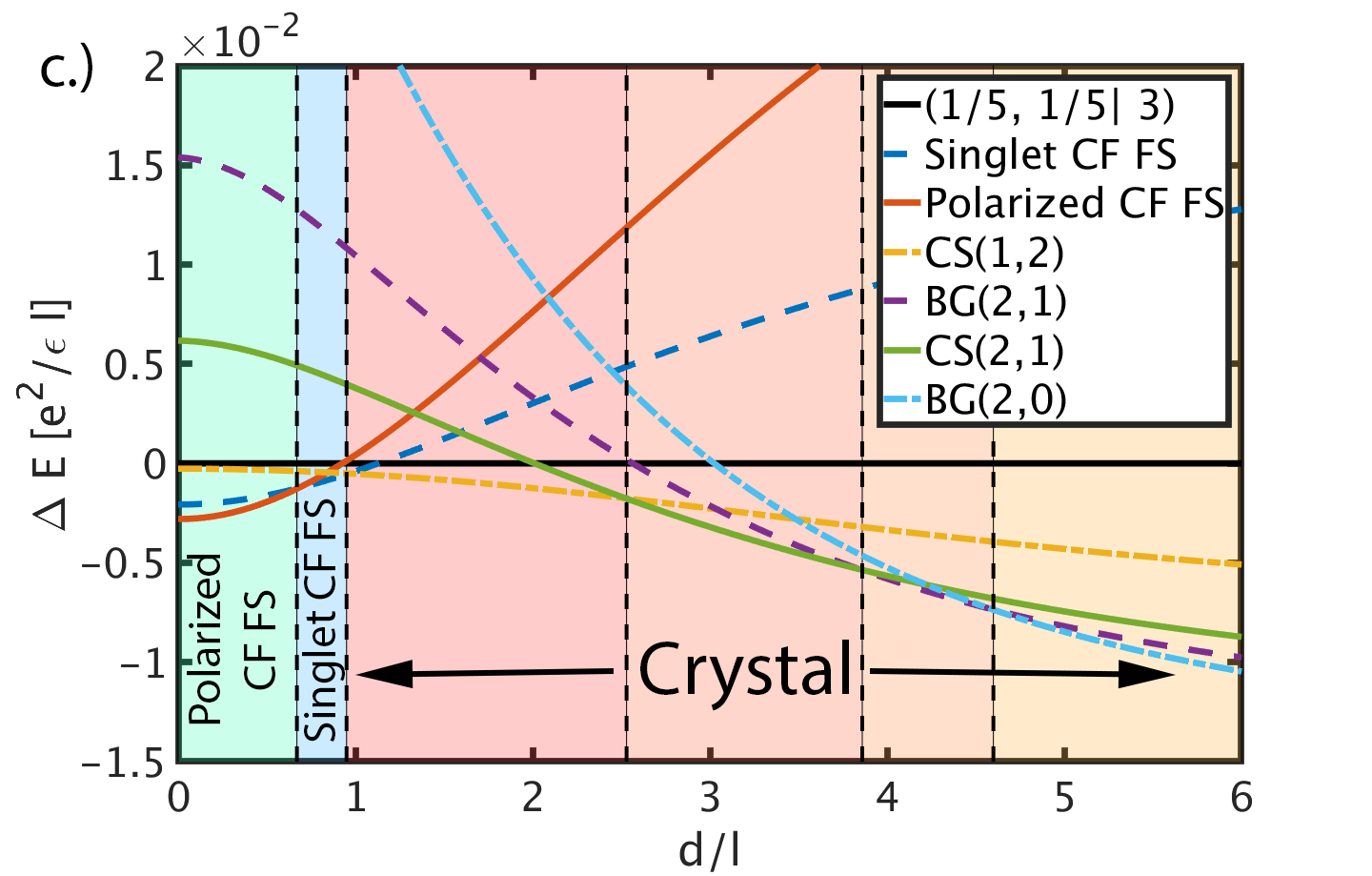}
\caption{\label{fig:60}Panel (a): Energies of several candidate states at $\nu=1/4$ as a function of density $\rho$ for a quantum well of width 60 nm. The different states are labeled as shown on the figure. All energies are thermodynamic values, measured relative to the energy of the Pfaffian state. Only the CFFS and 22111 states become ground states for the parameters studied. The inset shows the electron density as a function of transverse position for densities $1.5\times10^{11}~{\rm cm}^{-2}$ and $2.0\times10^{11}~{\rm cm}^{-2}$ as given by LDA for a quantum well of width 60 nm. Panel (b): The calculated phase diagram at $\nu=1/4$ as a function of the quantum well width and density. In the region of parameter space shown in the figure only the CFFS and 22111 states are realized. We also include experimental results, shown by black squares, taken from Refs.~\cite{Luhman08, Shabani09a}. Panel (c): Energies of several bilayer states as a function of the layer separation $d/l$. We have studied 11 liquid states (Table~\ref{tab:one_quarter_bilayer}) and 24 crystal states (SM). Here we omit the high energy states (see SM for more complete results) and show the energies of the $(1/5,1/5|3)$ state, the pseudospin singlet CFFS, the pseudospin polarized CFFS, and several crystal states (notation explained in text). All energies are measured relative to the $(1/5,1/5|3)$ state. No FQHE is stabilized.
}
\end{figure*}

We now come to FQHE at $\nu=1/4$ in WQWs. Which state occurs as the ground state is an energetic question. We consider the following candidate states:
\begin{equation}
\Psi^{\rm CFFS}= {\cal P}_{\rm LLL}\Phi^{{\rm Fermi~sea}} \Phi_1^{4}
\label{eq_CFFS_1_4}
\end{equation}
\begin{equation}
\Psi^{\rm Pf} = \text{Pf}\left(\frac{1}{z_i-z_j} \right)\Phi_1^{4} 
\label{eq_Pf_1_4}
\end{equation}
\begin{equation}
\Psi^{\rm{Pf}_{\textit{l}=3}} = \text{Pf}\left(\frac{1}{(z_i-z_j)^3} \right)\Phi_1^{4} 
\label{eq_Pf3_1_4}
\end{equation}
\begin{equation}
\label{eq:parton_JK_proj}
\Psi^{22111} = {\cal P}_{\rm LLL}\Phi_2\Phi_2\Phi_1^3 \sim \frac{[{\cal P}_{\rm LLL}\Phi_2\Phi_1^2]^2}{\Phi_1} = \frac{\Psi_{2/5}^2}{\Phi_1} 
\end{equation}
\begin{equation}
\Psi^{\bar{2}\bar{2}11111} = {\cal P}_{\rm LLL}[\Phi^{*}_2]^{2}\Phi_1^5 \sim [{\cal P}_{\rm LLL}\Phi_2^*\Phi_1^2]^2\Phi_1 = \Psi_{2/ 3}^2\Phi_1.
\end{equation}
These represent the compressible CF Fermi sea (CFFS)~\cite{Halperin93,Rezayi94}, the MR Pfaffian state~\cite{Moore91}, an $l=3$ pairing Pfaffian state, the 22111 state, and the $\bar{2}\bar{2}11111$ state. (We assume throughout this work that the magnetic field $B$ is large enough to freeze the spin degree of freedom.) The Pfaffian of an antisymmetric matrix $M_{i,j}$ is defined as ${\rm Pf}(M_{i,j})\sim \mathcal{A}(M_{1,2}M_{3,4}\cdots M_{N-1,N})$ with $\mathcal{A}$ representing antisymmetrization. The $22111$ and $\bar{2}\bar{2}11111$ states are projected into the LLL as shown above; this form allows the Jain-Kamilla projection~\cite{Jain97,Jain97b} to obtain the LLL states for up to 40 and 36 particles, respectively, in the spherical geometry~\cite{Haldane83} (see Supplemental Material (SM)~\cite{SM-Faugno-14} accompanying this paper). Particle-hole (PH) symmetry implies that the anti-Pfaffian state has the same energy as the Pfaffian state. We do not consider the so-called PH-Pfaffian state~\cite{Son15} because its wave function $\mathcal{P}_{\rm LLL}\text{Pf}(\frac{1}{z_i^*-z_j^*})\Phi_1^4$~\cite{Jolicoeur07,Zucker16,Mishmash18,Balram18} is not amenable to calculations for large systems, precluding a reliable thermodynamic limit for the energy.

The 22111 state is topologically distinct from the Pfaffian, anti-Pfaffian and the PH-Pfaffian states, which have chiral central charges of $c=3/2$, $-1/2$, and $1/2$, respectively. Nevertheless, while $\Psi^{\rm Pf}$ and $\Psi^{\rm{Pf}_{\textit{l}=3}}$ represent CF pairing in an obvious manner, the 22111 and $\bar{2}\bar{2}11111$ states are also paired states of composite fermions~\cite{Read00,Levin07,Lee07,Son15,Balram18}, which can be seen as follows. At filling fraction $\nu = 1/4$, a natural set of FQHE states to consider correspond to attaching $4$ vortices to each electron to obtain a composite fermion that sees zero magnetic field on average. In the parton construction this amounts to writing the electron operators as $\wp =  b \psi$, where $b$ is a boson that forms a $\nu_b = 1/4$ bosonic Laughlin FQHE state, while $\psi$ is the composite fermion. If we specialize to the case where $\psi$ forms a paired state, we can consider any odd $\ell$ angular momentum pairing. This leads to wave functions of the form
$\Psi^{\rm CF-paired}_\ell = \Psi^{\rm paired}_\ell \Phi_1^4$, where $\Psi^{\rm paired}_\ell$ is the wave function of an angular momentum $\ell$ paired superconductor of spinless fermions. $\Psi^{\rm CF-paired}_\ell$ describes a state with central charge $c = \ell/2 + 1$. The edge theory consists of $\ell$ chiral Majorana modes, together with a charge mode described by a single chiral boson. Because there is a unique topological quantum field theory with Ising quasiparticles for a given chiral central charge~\cite{Rowell09}, it follows that the  22111, the Pfaffian, the PH-Pfaffian and the anti-Pfaffian (or $\bar{2}\bar{2}11111$) states are topologically equivalent, respectively, to $\ell=3$, $1$, $-1$, and $-3$ paired states of composite fermions (Table~\ref{tab:one_quarter}). In particular, the 22111 state corresponds to an $f$-wave superconductivity of composite fermions~\cite{Balram18}. The 22111 state and $\Psi^{\rm{Pf}_{\textit{l}=3}}$ represent two different choices for the $f$-wave pair wave functions; while topologically equivalent, these two states are  microscopically very different, as seen below.

Since FQHE at $\nu=1/4$ is seen only in a WQW, it is crucial to incorporate into the calculation the variation in the interaction due to transverse wave function $\xi(x)$ of the electrons in a realistic fashion, where $x$ is the transverse coordinate. We determine $\xi(x)$ via the local density approximation (LDA)~\cite{Ortalano97} for a given width and electron density at zero magnetic field (see examples in Fig.~\ref{fig:60}a). This results in a modified effective interaction given by $V_{\rm eff}(r) = \int dx_1 \int dx_2 \frac{|\xi(x_1)|^2|\xi(x_2)|^2}{\sqrt{r^2+|x_1-x_2|^2}}$, where $r$ is the distance between the electrons within the plane. All energies are quoted in units of $e^2/\epsilon l$, where $\epsilon$ is the dielectric constant of the background host material and $l=\sqrt{\hbar c/eB}$ is the magnetic length.

The thermodynamic limits for the energies of various candidate states as a function of density are plotted in Fig.~\ref{fig:60}a for a quantum well of width 60 nm (see SM~\cite{SM-Faugno-14} for details). (The energy of $\Psi^{\rm{Pf}_{\textit{l}=3}}$ is much higher than that of other candidate states, typically by $0.1$ $e^2/\epsilon l$, and is not shown.) From similar calculations at other quantum well widths, we obtain the phase diagram presented in Fig.~\ref{fig:60}b. For small widths and small densities, the CFFS state dominates, but when the width and density are made large enough, the 22111 state becomes the ground state. The Pfaffian, anti-Pfaffian, or the $\bar{2}\bar{2}11111$ states are not realized in any part of the parameter space we have studied.  Fig.~\ref{fig:60}b also shows (solid squares), for two QW widths, the densities where the 1/4 FQHE has been first seen to appear in experiments~\cite{Luhman08, Shabani09a, Shabani09b}.

\begin{table}[b]
\begin{center}
\begin{tabular} { | c | c| c | c | c | c | }
\hline
State &  $\ell$ & shift $\mathcal{S}$ & $\alpha_{\rm e}$  &  $\alpha_{\rm qp}$ & central charge $c$\\
\hline
CFFS & -- &  4  & --  & --  & --\\
\hline
22111 &  $3$ & 7   & 9  & $-1/8$ & 5/2\\
\hline
Pfaffian$(l=3)$ &  $3$ & 7   & 9  & $-1/8$ & 5/2\\
\hline
Pfaffian &   $1$ & 5   & 9  & $-5/8$ & 3/2 \\
\hline
PH Pfaffian &  $-1$ & 3   & --  & -- & 1/2\\
\hline
$\bar{2}\bar{2}11111$ &  $-3$ & 1   & --  & -- & -1/2\\
\hline
\end{tabular}
\end{center}
\caption{\label{tab:one_quarter} This table gives the shift $\mathcal{S}$ on the sphere, the electron and quasiparticle tunneling exponents $\alpha_{\rm e}$ and $\alpha_{\rm qp}$ (defined so that the tunnel current behaves as $I\sim V^{\alpha}$), and the chiral central charge $c$ for several states at $\nu=1/4$. The central charge and the shift are related to the thermal Hall conductance and the Hall viscosity. Dashes indicate that the quantity is not expected to be quantized to a universal value due to the edge theory not being fully chiral or the bulk being gapless.}
\end{table}

\begin{table}[b]
\begin{center}
\begin{tabular} { | c | c |}
\hline
\multicolumn{2}{|c|}{Bilayer States at $\nu$ = 1/4}\\
\hline
State & wave function\\
\hline
$(1/8_{\rm CFFS},\ 1/8_{\rm CFFS}|\ 0)$ & $\Psi^{\rm CFFS}_{1/8}(z^\uparrow)\Psi^{\rm CFFS}_{1/8}(z^\downarrow)$\\
\hline
$(1/7,\ 1/7|\ 1)$ & $\Phi_1^7(z^\uparrow)\Phi_1^7(z^\downarrow)\Pi_{i,j}(z^\uparrow_i-z^\downarrow_j)$\\
\hline
$(1/6_{\rm CFFS},\ 1/6_{\rm CFFS}|\ 2)$ & $\Psi^{\rm CFFS}_{1/6}(z^\uparrow)\Psi^{\rm CFFS}_{1/6}(z^\downarrow)\Pi_{i,j}(z^\uparrow_i-z^\downarrow_j)^2$\\
\hline
$(1/5,\ 1/5|\ 3)$ & $\Phi_1^5(z^\uparrow)\Phi_1^5(z^\downarrow)\Pi_{i,j}(z^\uparrow_i-z^\downarrow_j)^3$\\
\hline
singlet CFFS & $\Psi^{\rm CFFS}_{1/4}(z^\uparrow)\Psi^{\rm CFFS}_{1/4}(z^\downarrow)\Pi_{i,j}(z^\uparrow_i-z^\downarrow_j)^4$\\
\hline
$(1/4_{\rm Pf},\ 1/4_{\rm Pf}| 4)$  & $\Psi^{\rm Pf}_{1/4}(z^\uparrow)\Psi^{\rm Pf}_{1/4}(z^\downarrow)\Pi_{i,j}(z^\uparrow_i-z^\downarrow_j)^4$\\
\hline
Pf $\times(1/6,\ 1/6|\ 2)$ & $\text{Pf}(\frac{1}{z_i-z_j})\Phi_1^6(z^\uparrow)\Phi_1^6(z^\downarrow)\Pi_{i,j}(z^\uparrow_i-z^\downarrow_j)^2$\\
\hline
$(1/6_{\rm Pf},\ 1/6_{\rm Pf}|\ 2)$ & $\Psi^{\rm Pf}_{1/6}(z^\uparrow)\Psi^{\rm Pf}_{1/6}(z^\downarrow)\Pi_{i,j}(z^\uparrow_i-z^\downarrow_j)^2$\\
\hline
polarized CFFS & $\Psi^{\rm CFFS}(z^\uparrow,z^\downarrow)$\\
\hline
polarized Pf  & $\Psi^{\rm Pf}(z^\uparrow,z^\downarrow)$\\
\hline
singlet $2_{_{\uparrow\downarrow}}2111$ & $\mathcal{P}_{\rm LLL}\Phi_1(z^\uparrow)\Phi_1(z^\downarrow)\Phi_2(z)\Phi_1^3(z)$\\
\hline
\end{tabular}
\end{center}
\caption{\label{tab:one_quarter_bilayer} Candidate liquid state wave functions at $\nu=1/4$ in a bilayer system. The coordinates $z^\uparrow$ and $z^\downarrow$ denote different layers, while $z$ denotes all coordinates. The terms singlet and polarized refer to ``layer polarization."}
\end{table}

A sufficiently wide quantum well can behave as a bilayer, which raises the question whether a two-component FQHE state could also be competitive~\cite{Papic09}. The following considerations point to a single component state. (i) The experimental onset of the 1/4 FQHE with increasing width or density agrees well with the phase boundary obtained in our single component calculation (Fig.~\ref{fig:60}a). (ii) The competition between one and two-component states depends sensitively on the gap $\Delta_{\rm SAS}$ separating the symmetric and the antisymmetric bands. A large $\Delta_{\rm SAS}$ favors a one-component state.  From the LDA calculation, the value of $\Delta_{\rm SAS}$ at the phase boundary in Fig.~\ref{fig:60}b is $\sim$ 0.1,  0.08 and 0.06 $e^2/\epsilon l$, respectively, for QWs of widths 50 nm, 60 nm and 70 nm.  While seemingly small, $\Delta_{\rm SAS}$ is large compared to typical Coulomb energy differences between competing states (e.g. the Coulomb energy differences are $<$0.005 $e^2/\epsilon l$ in  Fig.~\ref{fig:60}a). For another two component system, namely spinful electrons in a single layer, the system in the vicinity of $\nu=1/2$ becomes fully polarized (i.e. single-component) when $E_{\rm Z}  \gtrsim 0.01 e^2/\epsilon l$ for WQWs~\cite{Liu14,Zhang16}, where the Zeeman splitting $E_{\rm Z}$ is analogous to $\Delta_{\rm SAS}$. It is therefore likely that two-component states are not relevant for $\Delta_{\rm SAS} \sim 0.05-0.10 \; e^2/\epsilon l$. (iii) In the vicinity of $\nu=1/4$, the FQHE states and the CFFS of spinful electrons are predicted to be single-component (i.e. fully spin polarized) even for $E_{\rm Z}=0$~\cite{Park99,Balram15a}. (iv) Finally, we have considered an ideal bilayer system of two two-dimensional systems separated by a distance $d$. We have studied a total of 11 compressible and incompressible liquid states (Table~\ref{tab:one_quarter_bilayer}) and 24 different crystal states (SM). The crystal labeled BG($2p,m$) refers to a ``Bilayer Graphene" crystal of composite fermions with $2p$ attached vortices, with $m$ interlayer zeros; CS($2p,m$) refers to an analogous ``Correlated Square" crystal~\cite{Faugno18}. For $d=0$ the system is formally equivalent to that of spinful particles in a single layer with zero Zeeman energy. Here, as mentioned above, the ground state is a fully pseudospin polarized CFFS, which has lower exchange energy than the pseudospin singlet CFFS because of exchange effects. We find, unexpectedly, that as $d$ is increased, a transition occurs into a pseudospin singlet CFFS, which is followed by a sequence of correlated CF crystals at larger $d/l$ (see Fig.~\ref{fig:60}c). We thus predict that no FQHE will occur at $\nu=1/4$ in a bilayer system. This is consistent with current experiments in GaAs double QW systems~\cite{Eisenstein92}, and can be tested more accurately in double layer graphene where a plethora of FQHE states have recently been observed~\cite{Kim18b,Dean18}. 

These considerations make it plausible that the single-component 22111 state is stabilized in WQWs. Nonetheless, a decisive confirmation requires further experimental evidence, and in the remainder of this paper we outline certain experimental consequences of the 22111 state. 

The thermal Hall conductance of a FQHE sate is given by $c[\pi^2k_B^2 /(3h)]T$, where $c$ is the central charge~\cite{Kane97}. It can thus decisively distinguish between different candidate states at $\nu=1/4$, as they have different values of $c$ (see Table~\ref{tab:one_quarter}). An advantage of thermal Hall conductance is that it is robust against edge reconstruction.

We next come to tunneling exponents. We first consider quasiparticle tunneling at a quantum point contact (QPC) separating two edges of the \it same \rm quantum Hall fluid. (See SM~\cite{SM-Faugno-14} for the properties of various quasiparticles.) This tunneling is expected to be dominated by the minimally charged quasiparticle carrying charge $1/8$, given by the operator $\sigma e^{i\varphi/2\sqrt{\nu^{-1}}}$. For $\ell = 1$, $\sigma$ is the usual Ising spin field; for general $\ell$, it is the primary field that changes the sign of the boundary condition of each of the chiral Majorana fermions and has scaling dimension $\ell/16$. The chiral boson $\varphi$ carries the charge. For $\ell > 0$, the quasiparticle operator has scaling dimensions $(h,\bar{h}) = (\frac{\nu}{8} + \frac{\ell}{16},0)$, where $h$ and $\bar{h}$ are the left and right scaling dimensions. This implies that at a QPC, the backscattering tunneling current would be (with $\nu=1/4$)~\cite{Wen92b}
\begin{align}
I \propto V^{4(h+\bar{h}) - 1} = V^{(\ell+2)/4 - 1} = V^{(2\ell-7)/8}.
\end{align}
For $\ell > 0$, since the edge theory is fully chiral, these exponents are quantized and universal due to charge $1/8$ quasiparticles (in the absence of edge reconstruction).  For $\ell < 0$, this operator has scaling dimensions $(h,\bar{h}) = (\frac{\nu}{8},\frac{\ell}{16})$.  In the unperturbed edge theory, we would therefore expect $I \propto V^{4(h+\bar{h}) - 1} = V^{(2|\ell|-7)/8}$. However, since the theory is not fully chiral there are marginal perturbations of the edge theory that can modify the scaling dimensions. Therefore for $\ell < 0$ we do not expect these exponents to be universal and thus not quantized. In particular, we can consider perturbations $\delta L = i \alpha_{ij} \partial \varphi \eta_i \partial \eta_j$, for coupling constants $\alpha_{ij}$, where $\eta_i$ for $i = 1, \cdots, \ell$ are the chiral Majorana fermions. These perturbations are marginal, having scaling dimension two, and can change the exponents of the quasiparticle and the electron operators. 

We next consider tunneling of an electron between two distinct adjacent FQHE fluids. The edge theory has $\ell$ chiral Majorana fermions, $\eta_i$, for $i = 1,\cdots, \ell$. We therefore have $\ell$ different types of electron operators: $\Psi_{e;i} \propto \eta_i e^{i \sqrt{\nu^{-1}} \varphi}$, and in general can consider 
a linear combination of the above operators:
$\Psi_e = e^{i \sqrt{\nu^{-1}} \varphi} \sum_{i = 1}^\ell a_i \eta_i + \cdots$
where the $a_i$ are some constant coefficients for the expansion of the electron in terms of long wavelength field operators, and $\cdots$ indicate higher order (less relevant) operators in the expansion.
For $\ell > 0$, this operator has scaling dimensions
$(h,\bar{h}) = (\frac{1}{2\nu} + \frac{1}{2},0)$, where 
where $h$ and $\bar{h}$ are the left and right scaling dimensions. The tunneling current behaves as 
\begin{align}
I \propto V^{4(h+\bar{h}) - 1} = V^{9}
\end{align}
For $l < 0$, this operator has scaling dimensions $(h,\bar{h}) = (\frac{1}{2\nu},\frac{1}{2})$.  While naively we would still get the same tunneling exponent, namely $I \propto V^{4(h+\bar{h}) - 1} = V^{9}$. As before, for $\ell>0$ the exponent is quantized and universal (assuming no reconstruction) but not for $l<0$.

One can similarly consider tunneling of an electron from an external Fermi liquid~\cite{Chang96,Grayson98,Chang03}. In this case the tunneling current
becomes $I \propto V^{2(h+\bar{h}) } = V^5$. 

Finally, we note that the Hall viscosity is conjectured to be quantized at $\eta_{\rm H}=\hbar \rho {\cal S}/4$, where ${\cal S}$ is the ``shift" in the spherical geometry~\cite{Read09} and $\rho$ is the density. The shifts for different candidate states are shown in Table~\ref{tab:one_quarter}.

In summary, we have presented extensive calculations that make a strong case that the $\nu=1/4$ FQHE in wide quantum wells is the realization of a new kind of single-component non-Abelian state that is topologically equivalent to $f$-wave pairing of composite fermions. We have listed many experimental consequences of this state. If confirmed, it will provide a convenient new platform for creating and studying non-Abelian anyons.

\begin{acknowledgments}
The work at Penn State was supported by the U. S. Department of Energy, Office of Basic Energy Sciences, under Grant no. DE-SC0005042. The Center for Quantum Devices is funded by the Danish National Research Foundation. This work was supported by the European Research Council (ERC) under the European Union Horizon 2020 Research and Innovation Programme, Grant Agreement No. 678862 and the Villum Foundation. MB is supported by NSF CAREER (DMR-1753240), JQI-PFC-UMD and an Alfred P. Sloan Research Fellowship. Some portions of this research were conducted with Advanced CyberInfrastructure computational resources provided by The Institute for CyberScience at The Pennsylvania State University. 
\end{acknowledgments}

\appendix
\section{Supplemental Material}
% for ``Prediction of a non-Abelian fractional quantum Hall state with $f$-wave pairing of composite fermions in wide quantum wells''

In this Supplemental Material (SM), we (i) provide a primer on the spherical geometry, (ii) consider the case of asymmetric charge distributions, (iii) discuss details of the thermodynamic extrapolation of the energies of the various candidate states, (iv) describe other bilayer states that we considered in our work and (v) elucidate the topological properties of the $22111$ parton state derived from its effective edge theory. 

\section{Spherical Geometry}
\label{app:spherical}
All our calculations are done in the spherical geometry~\cite{Haldane83} in which $N$ electrons are confined to a spherical shell and subjected to a radial magnetic flux of $2Q(h/e)$ ($2Q$ is an integer) generated by a magnetic monopole, placed at the center of the sphere. It is convenient to introduce spinor coordinates $u=\cos(\theta/2)e^{i\phi/2}$ and $v=\sin(\theta/2)e^{-i\phi/2}$, where $\theta$ and $\phi$ are the polar and azimuthal angles on the sphere. In terms of these spinor coordinates, the chord distance between the $i^{\rm th}$ and $j^{\rm th}$ electron is $\sqrt{Q}l|u_iv_j-u_jv_i|$. The spherical geometry version of the Jastrow factor and the Pfaffian are obtained by replacing $(z_i-z_j)$ by $(u_iv_j-u_jv_i)$.

Many of our wave functions include the Slater determinant, $\Phi_n$, of $n$-filled Landau levels of electrons. The constituent single particle wave functions of $\Phi_n$ are given by the monopole harmonics $Y_{Q,l,l_{z}}$, with orbital angular momentum $l=|Q|+n$ and $l_{z}$ the $z$-component of the orbital angular momentum, $l_{z}=-l,-l+1,\cdots,l-1,l$. Therefore the degeneracy of the LL indexed by $n$ is $2(|Q|+n)+1$. Integer quantum Hall states filling the lowest $n$ LLs occur when $2Q=N/n-n$.

An incompressible FQH state at filling $\nu = n/(2pn+1)$ in the LLL occurs when $N$ composite experience an effective magnetic field generated by a monopole of strength $2Q^*=N/n-n$. The net magnetic monopole strength seen by the electrons $2Q = 2Q^* + 2p(N-1)\equiv \nu^{-1}N-\mathcal{S}$. Thus, the shift~\cite{Wen92} for these states $\mathcal{S}=n+2p$. Note that the filling factor in the spherical geometry is defined as $\nu=\lim_{N\rightarrow \infty}(N/2Q)$. The filling factor and shifts for parton wave functions can be found similarly to be
\begin{eqnarray}
\nu &=& \lim_{N\rightarrow \infty}\frac{N}{2Q} = \left[\sum_\lambda n_\lambda^{-1}\right]^{-1} \\
\mathcal{S} &=& \nu^{-1}N-2Q= \sum_\lambda n_\lambda .
\end{eqnarray}

\section{Asymmetric Charge Distribution in a Quantum Well}
\label{app:asym}
\begin{figure}[ht]
\includegraphics[width=3.5in]{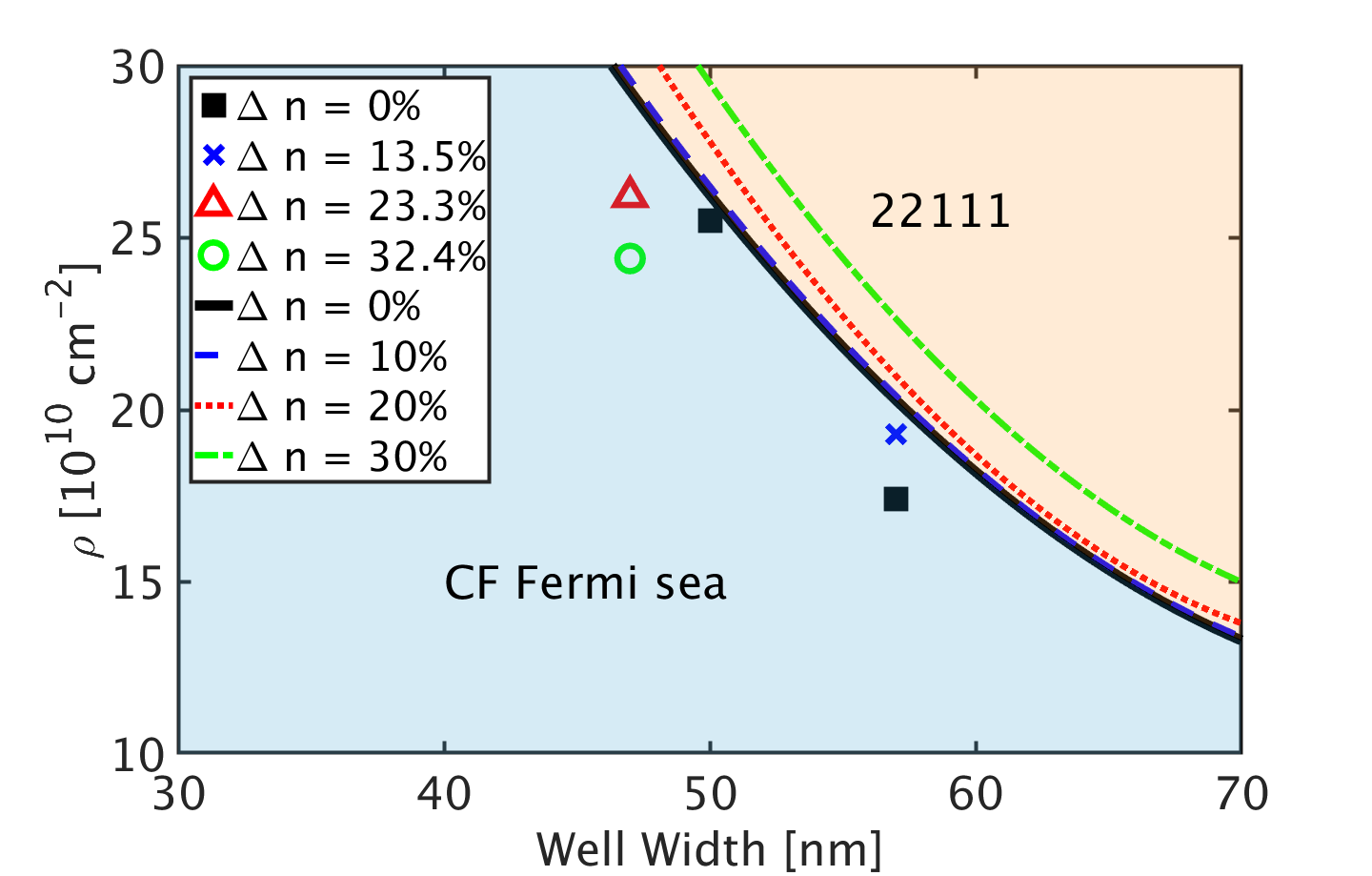}
\caption{\label{fig:asym}Phase diagram as a function of electron density and quantum well width with asymmetric charge distribution results. In the region of parameter space shown in the figure only the CFFS and 22111 states are realized.  We show the phase boundary between them for both symmetric and asymmetric charge distributions. We also include several experimental results, shown by symbols. Experimental results for symmetric quantum wells (black squares) are from Refs.~ \cite{Luhman08, Shabani09a}, and those for asymmetric quantum wells from Refs.~ \cite{Shabani09a, Shabani09b}. $\Delta n$ is obtained by taking the difference between the charge accumulated in each half of the well divided by the total charge.}
\end{figure}

All the results discussed in the main text are for quantum wells with a symmetric charge distribution. There have been additional experiments that observed an FQHE at 1/4 when the charge distribution is made asymmetric by applying a bias voltage~\cite{Shabani09a,Shabani09b}. We have constructed effective interactions for an asymmetric well and studied the phase diagram as a function of the density and well width at $1/4$. The asymmetry, $\Delta n$, is given as a percentage calculated by dividing the difference between the charges in the right and left halves of the well by the total charge. We report results for the symmetric case along side asymmetries of 10$\%$, 20$\%$, and 30$\%$ as shown in Fig.~\ref{fig:asym}. 

The asymmetric case seems to push the phase boundaries to higher well widths and densities. This is likely due to more electrons being concentrated into a smaller area, and thus appearing to be in a narrower quantum well. The theoretical phase boundary agrees reasonably well with the experimental observations. On the other hand, the asymmetric charge distribution results appear to agree less than in the symmetric case.

\section{Extrapolation to the thermodynamic limit for single-component states}
\label{app:extrap}
\begin{figure}[ht]
\includegraphics[width =3.5in]{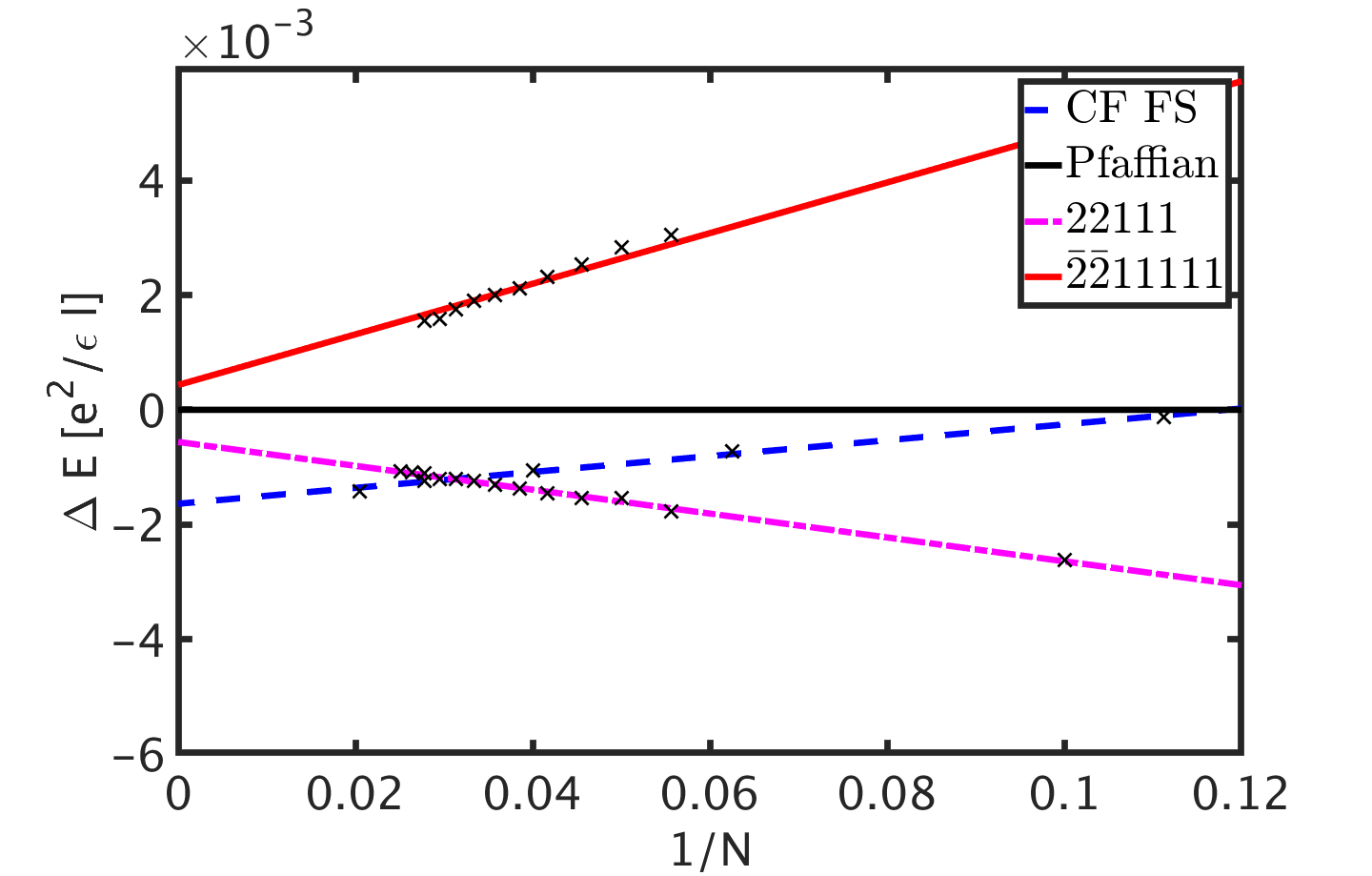}
\caption{Thermodynamic Limits for candidate states at $\nu=1/4$ for a quantum well width $w=60$ nm and electron density of $\rho=1.0\times 10^{11}$~cm$^{-2}$. We evaluate the energy difference between each candidate state and the Pfaffian for various system sizes and fit to $a/N+b$. The resulting value of $b$ is the energy difference of the state from the Pfaffian in the thermodynamic limit.}\label{fig:th10}
\end{figure}

\begin{figure}[ht]
\includegraphics[width =3.5in]{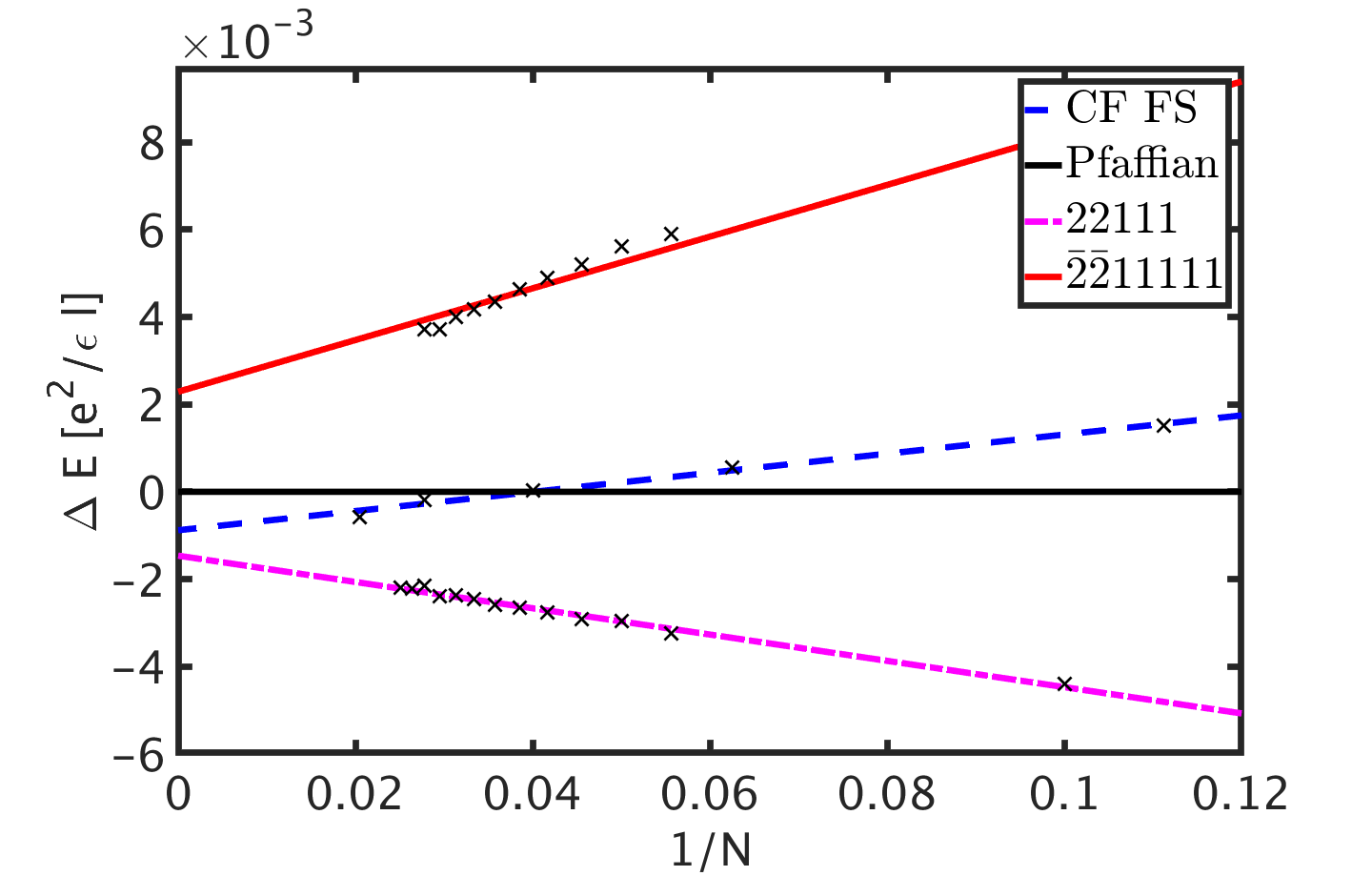}
\caption{Thermodnyamic Limits for candidate states at $\nu=1/4$ for a quantum well width $w=60$ nm and electron density $\rho=2.5\times 10^{11}$~cm$^{-2}$. We evaluate the energy difference between each candidate state and the Pfaffian for various system sizes and fit to $a/N+b$. The resulting value of $b$ is the energy difference of the state from the Pfaffian in the thermodynamic limit.}\label{fig:th25}
\end{figure}

To obtain the thermodynamic limit of the per-particle energy for each state, we extrapolate the finite system results as a linear function of $1/N$ to the limit $1/N\rightarrow 0$. In Figs.~\ref{fig:th10} and ~\ref{fig:th25} we show thermodynamic extrapolations for single component states in a quantum well of width 60 nm and electron densities 10$\times 10^{10}$~cm$^{-2}$ and 25$\times 10^{10}$~cm$^{-2}$. Before taking the thermodynamic limit, we multiply the energy by $\sqrt{2Q\nu/N}$ to correct for differences between the density in the finite system and the density in the thermodynamic limit~\cite{Morf86}. Finally, we subtract off the energy of the Pfaffian for single component states. Taking this difference suppresses the effects of background energy and provides a better linear fit for the energy. Since the Pfaffian is only defined for an even number of particles, to obtain the energy for an odd number of particles we use the method of interpolation.

\section{Bilayer States}
\label{app:bilayer}
To model a bilayer system, we consider a pair of two dimensional electron gases (2DEGs) separated by a distance $d$ measured in units of the magnetic length $l$. We introduce a pseudospin with two components, up ($\uparrow$) and down ($\downarrow$), where up and down represent the different layers. The intra- and inter-layer interactions are then given by
\begin{equation}
V_{\uparrow\uparrow}(r)=V_{\downarrow\downarrow}(r) = \frac{e^2}{\epsilon l}\frac{1}{r}
\end{equation}
\begin{equation}
V_{\uparrow\downarrow}(r) = \frac{e^2}{\epsilon l}\frac{1}{\sqrt{r^2+(d/l)^2}},
\end{equation}
where $r$ is measured in units of $l$. Since the interaction now depends on the pseudospin index, we consider a broader class of wave functions including the ones which are not an eigenstate of the total pseudospin operator. These wave functions include the Halperin states, their CF generalizations, several non-Abelian liquid states, and a large family of crystal states.

The two-component Halperin states are described by the wave function~\cite{Halperin84}
\begin{equation}
\Psi_{(n_1\ n_2| m)} = \Phi_1^{n_1}(z^\uparrow)\Phi_1^{n_2}(z^\downarrow)\Pi_{i,j}(z^\uparrow_i-z^\downarrow_j)^m,
\end{equation}
where $z^\uparrow$ and $z^\downarrow$ correspond to the coordinates of particles in different layers. The exponents $n_1$ and $n_2$ are odd integers while $m$ is any integer. Viewing these states as the product of two Laughlin states with an interlayer correlation term whose strength is determined by $m$, a natural generalization to a broader class of bilayer wave functions is given by~\cite{Jain07}
\begin{equation}
\Psi_{(\bar{\nu}_1\ \bar{\nu}_2|\ m)} = \Psi_{\bar{\nu}_1}\Psi_{\bar{\nu}_2}\Pi_{i,j}(z^\uparrow_i-z^\downarrow_j)^m.
\end{equation}
Here $\Psi_\nu$ is the CF wave function at filling $\nu$ and $\bar{\nu}_1$ and $\bar{\nu}_2$ are the filling factor of the two individual layers. We specialize to the case where $\bar{\nu}_1 = \bar{\nu}_2 = \bar{\nu}$. The single layer filling factor is related to the total filling factor $\nu_T$ by
\begin{equation}
\nu_T = \frac{2\bar{\nu}}{1+m\bar{\nu}}.
\end{equation}
From these we can obtain several liquid bilayer states at $\nu_T=1/4$ as shown in Table I of the main text.

Additionally, we consider a bilayer parton state, the so called 2$_{_{\uparrow\downarrow}}$2111. This is the singlet version of the fully polarized 22111 parton state considered in the main text. The wave function of this state is~(see Ref.~\cite{Belkhir93} for an analogous construction at $\nu=1/2$)
\begin{eqnarray}
\Psi^{2_{\uparrow\downarrow}2111} &=& \mathcal{P}_{\rm LLL}\Phi_1(z^\uparrow)\Phi_1(z^\downarrow)\Phi_2(z)\Phi_1^3(z) \nonumber \\
&\sim&\Phi_1(z^\uparrow)\Phi_1(z^\downarrow)\Psi_{2/5}(z)\Phi_1(z), 
\end{eqnarray}
where $z$ denotes the set of collective coordinates including all the $z^\uparrow$'s and $z^\downarrow$'s.

\begin{figure}[ht]
\includegraphics[width=3.5in]{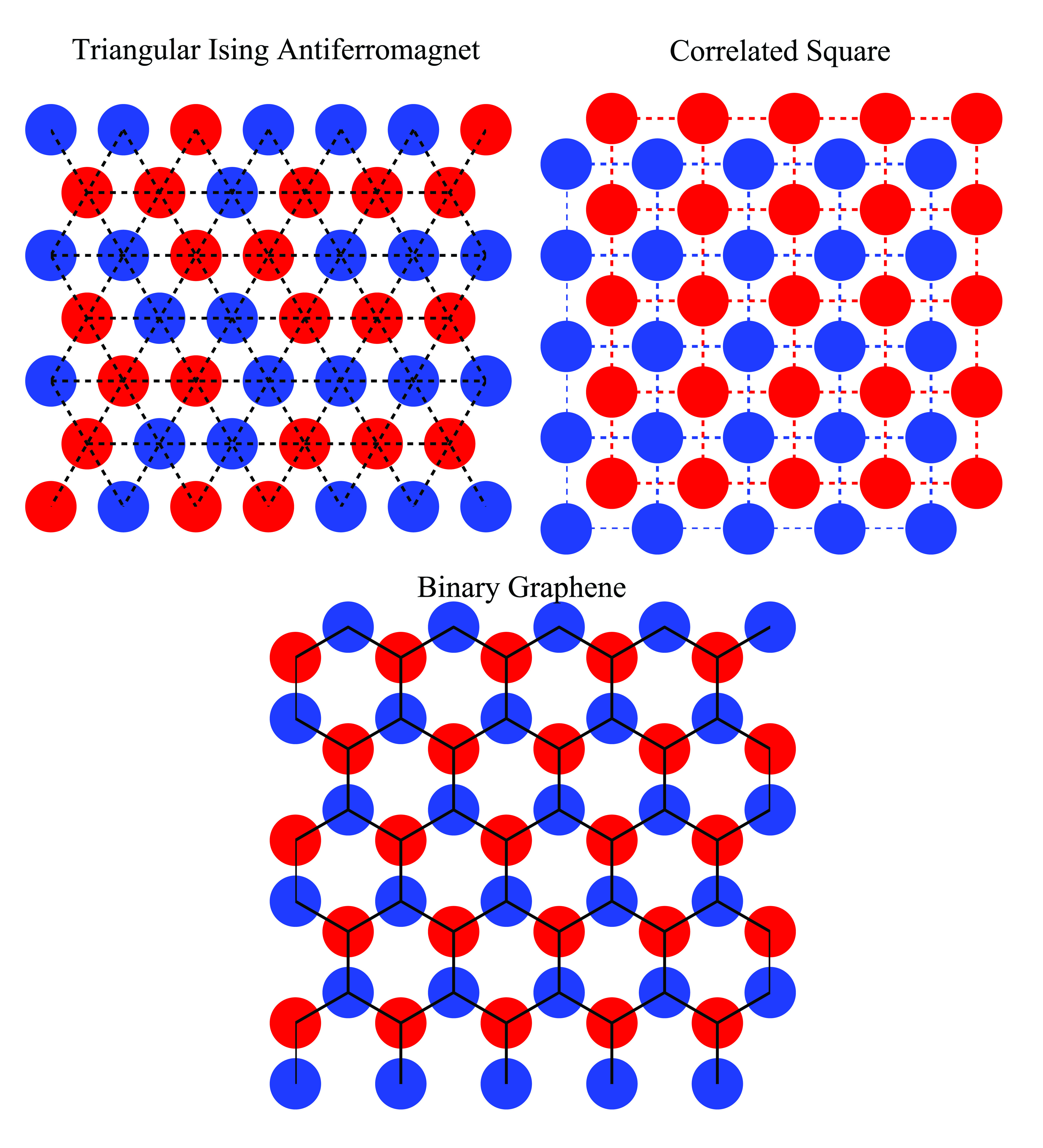}
\caption{2D representations of the bilayer crystal configurations. Blue and red denote particles in different layers.}
\label{fig:2Drep}
\end{figure}

Finally, we consider a large number of CF and electron crystals. The bilayer crystal wave function can be written as
\begin{eqnarray}
\Psi^{(2p,m)}_\nu &=& \det\left(\delta_p(z^\uparrow_i-R^{(\uparrow)}_j)\right)\Phi_1^{2p}(z^\uparrow)\det\left(\delta_p(z^\downarrow_i-R^{(\downarrow)}_j)\right) \nonumber \\
&\times &\Phi_1^{2p}(z^\downarrow)\Pi_{i,j}(z^\uparrow_i-z^\downarrow_j)^m.
\end{eqnarray}
Here $\delta_p(x)$ is the Dirac delta function projected into the LLL. In spherical geometry, the projected Dirac delta function is given by~\cite{Jain07}
\begin{equation}
\delta_p(z_i-R_j) = (U_j^*u_i+V_j^*v_i)^{2Q^*},
\end{equation}
where $U_j$ and $V_j$ are spinor coordinates corresponding to the crystal centers and $u_i$ and $v_i$ are spinor coordinates for the particles.
The coordinates $R^{(\uparrow)}_j$ and $R^{(\downarrow)}_j$ are the crystal centers in opposite layers as determined by minimizing the Coulomb energy of classical point charges confined to the surface of a sphere. Previous studies of bilayer FQH systems have found that there are three likely crystal configurations, triangular Ising antiferromagnetic (TIAF), correlated square (CS), and binary graphene (BG)~\cite{Faugno18}. TIAF is a triangular crystal when viewed from above in which every triangle consists of at least one member of both layers. CS consists of a square lattice in each layer, staggered so that the sites in one layer lie across from the empty centers in the other. BG is a graphene like crystal when viewed from above where each sublattice resides entirely in one layer. (See Fig.~\ref{fig:2Drep} for a schematic view of these three crystals.) We enumerate a total of 24 crystal states at filling factor 1/4 by choosing all possible values of $2p$ and $m$, such that $2Q^* = 8(N-1) - 2p(N-1) - mN > N$ (to ensure that $\nu^{*}=\lim_{N\rightarrow\infty}N/(2Q^{*})<1$, i.e., composite fermions occupy only the lowest $\Lambda$ level, which is necessary for constructing a crystal). 

\begin{figure}[ht]
\includegraphics[width=3.5in]{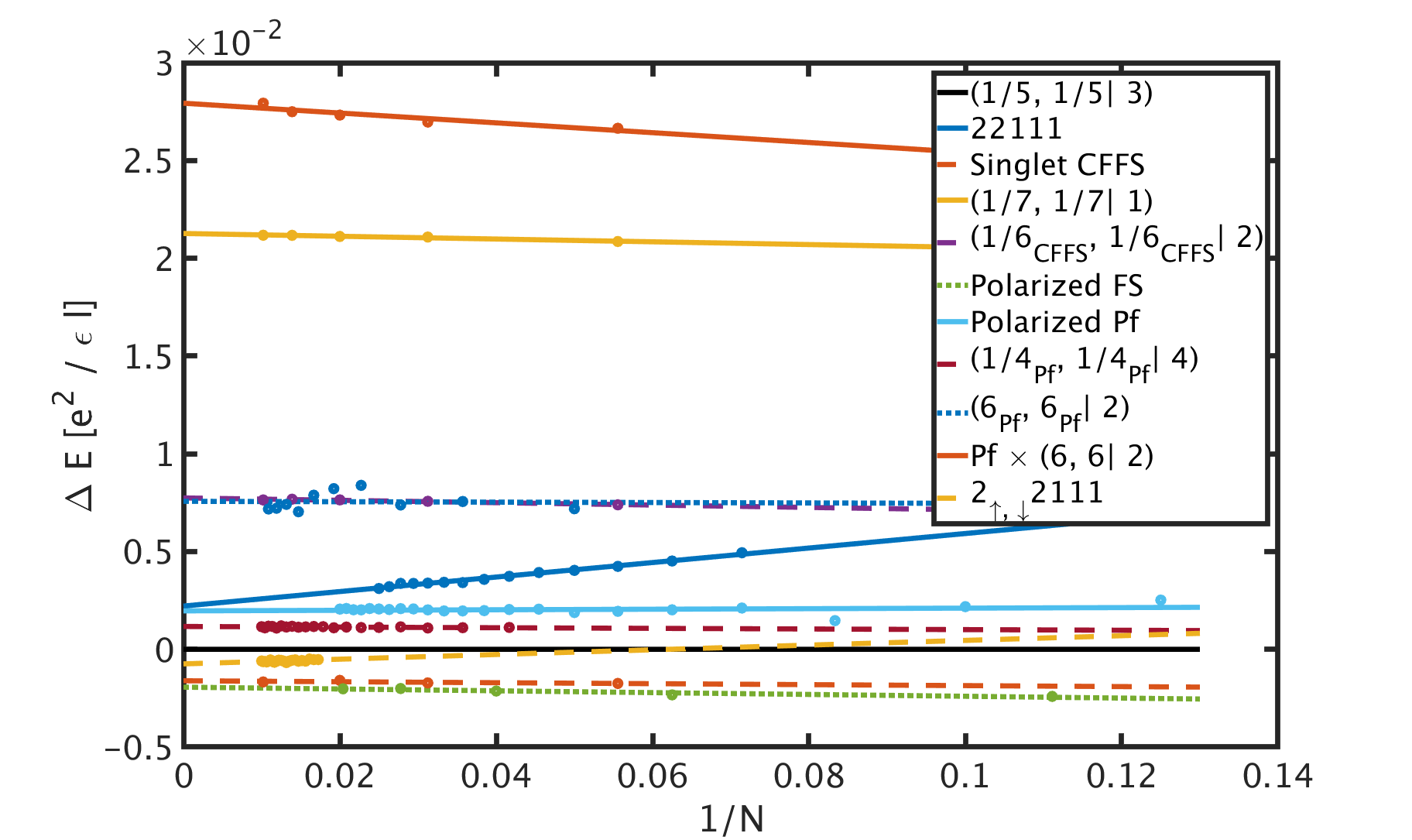}
\caption{Thermodynamic limit of the energy difference of bilayer liquid states from the $(1/5,~1/5|\ 3)$ state for a layer separation of $d/l = 0.5$.}
\label{fig:bliqTL}
\end{figure}

\begin{figure}
\includegraphics[width=3.5in]{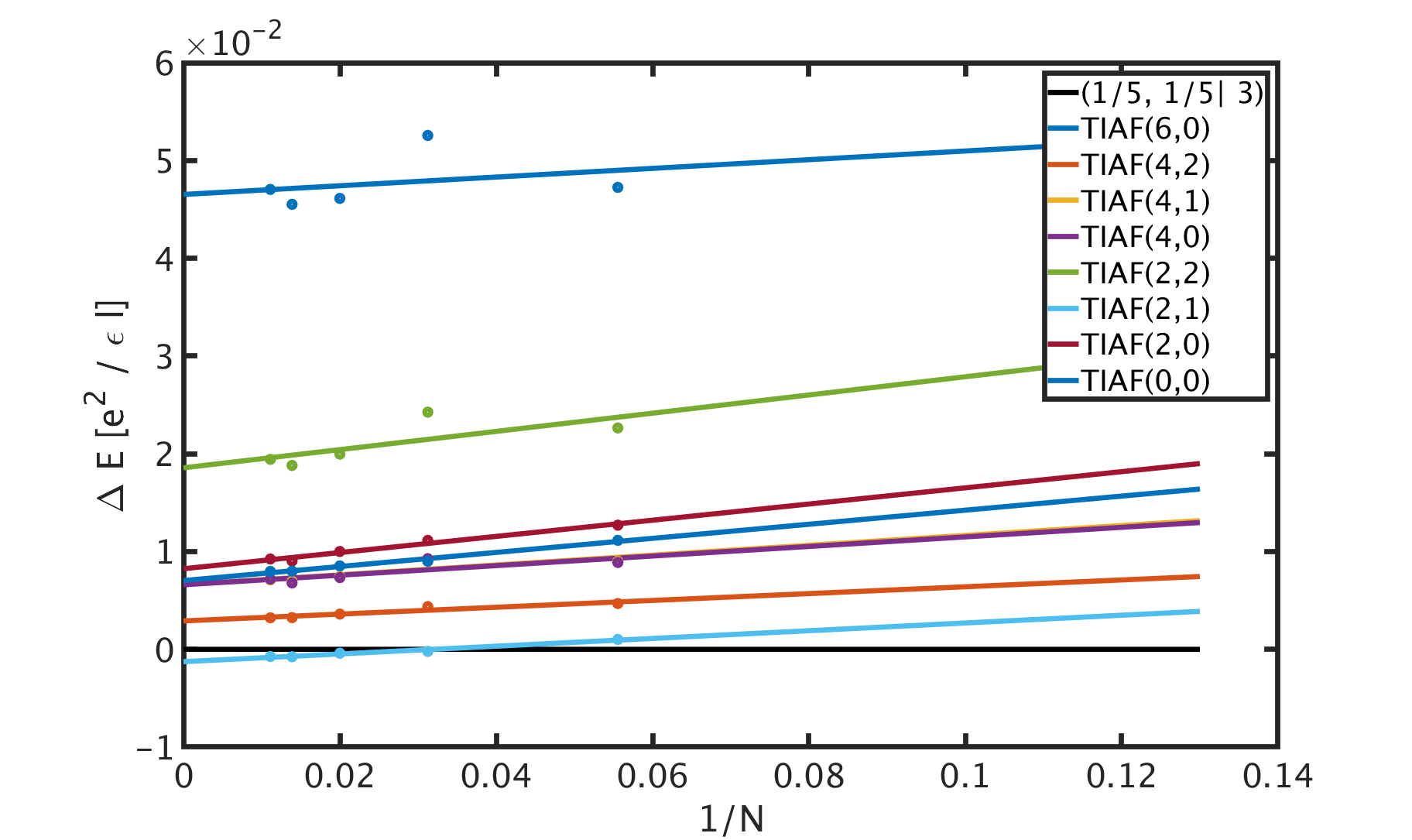}
\includegraphics[width=3.5in]{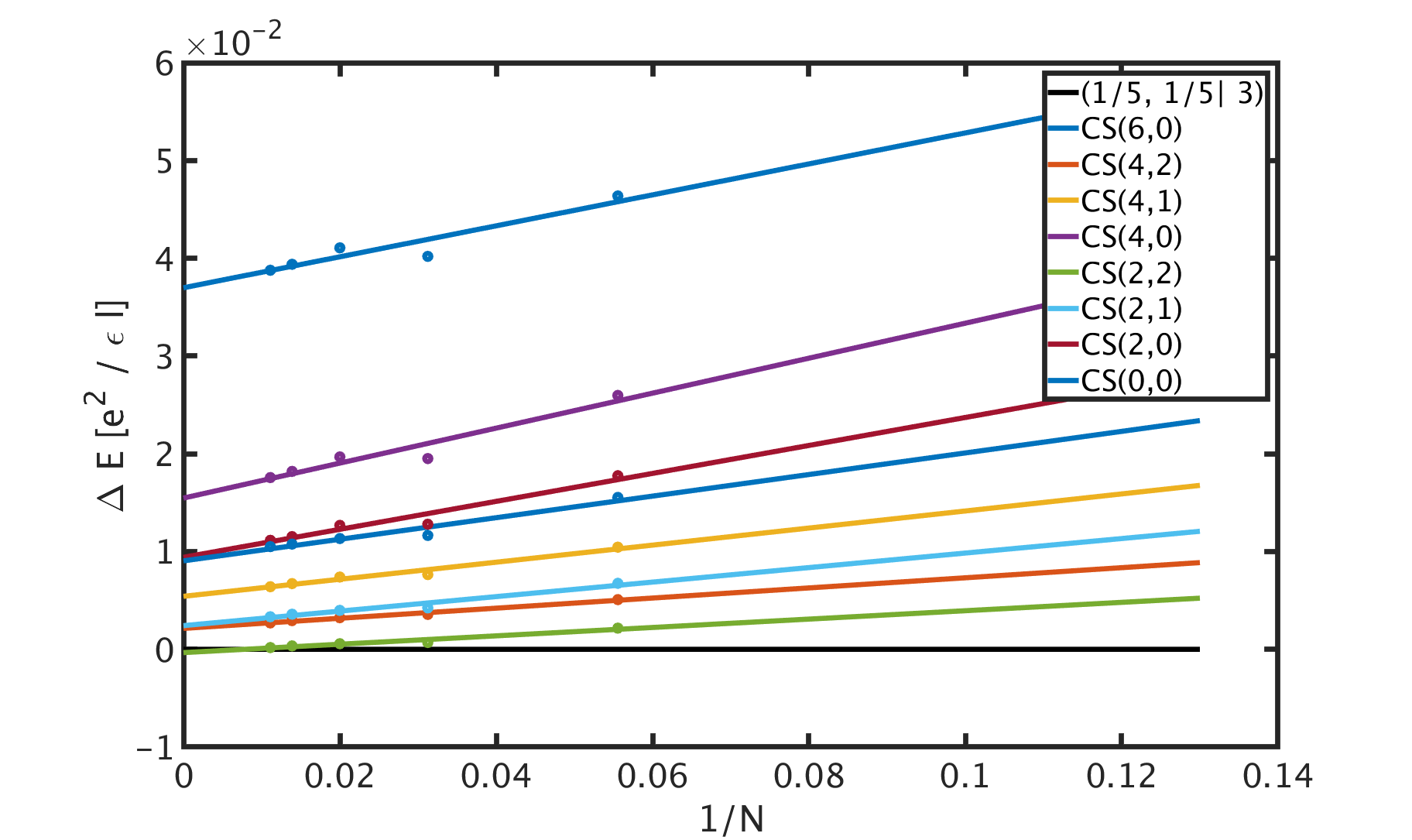}
\includegraphics[width=3.5in]{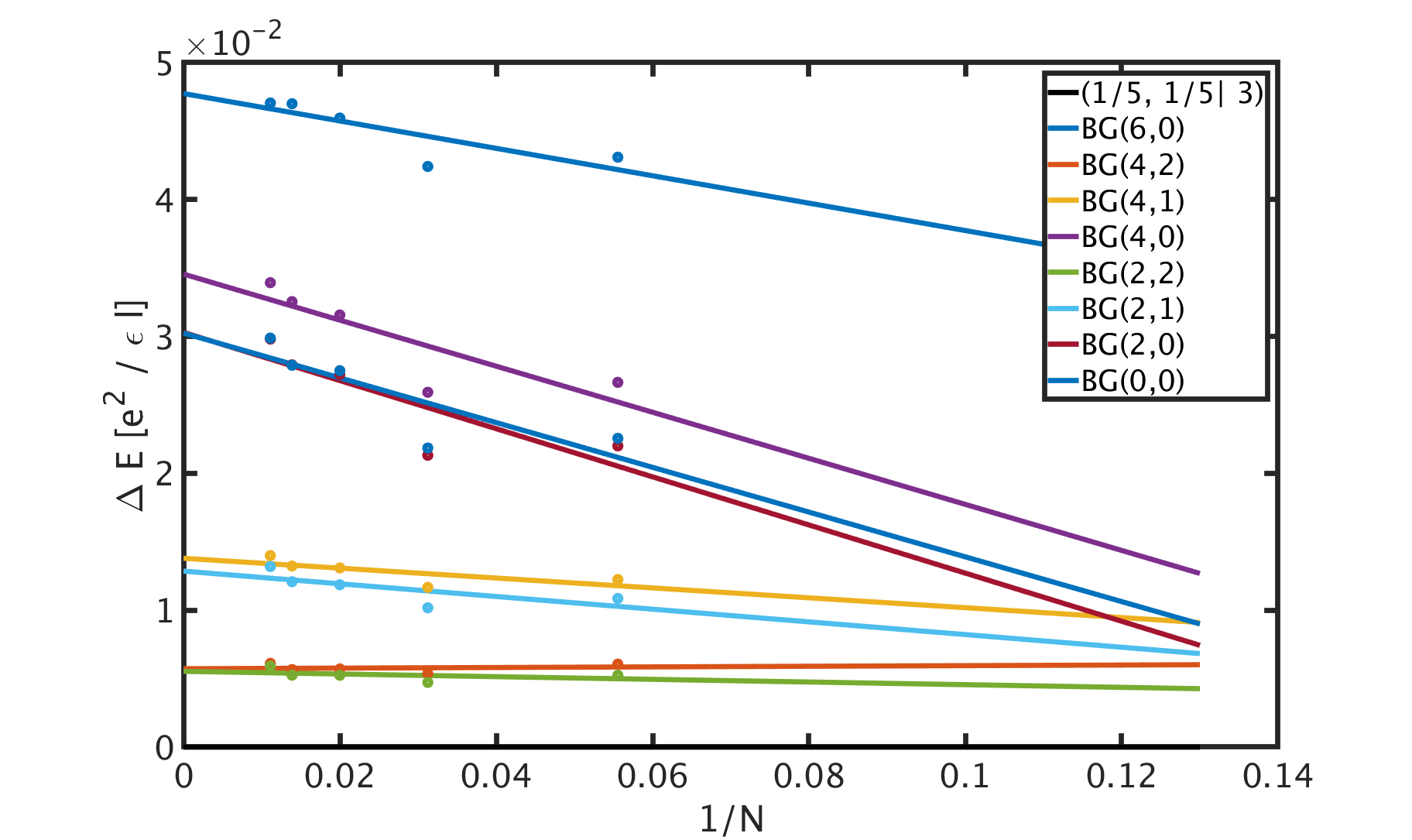}
\caption{Thermodynamic limit of the energy difference of TIAF (upper panel), CS (middle panel) and BG (bottom panel) crystal states from the $(1/5,~1/5|3)$ state for a layer separation of $d/l = 0.5$. The legend gives the values of $(2p,m)$ corresponding to each crystal state.}
\label{fig:bilayer_crystalsTL}
\end{figure}

\begin{figure}[ht]
\includegraphics[width=3.5in]{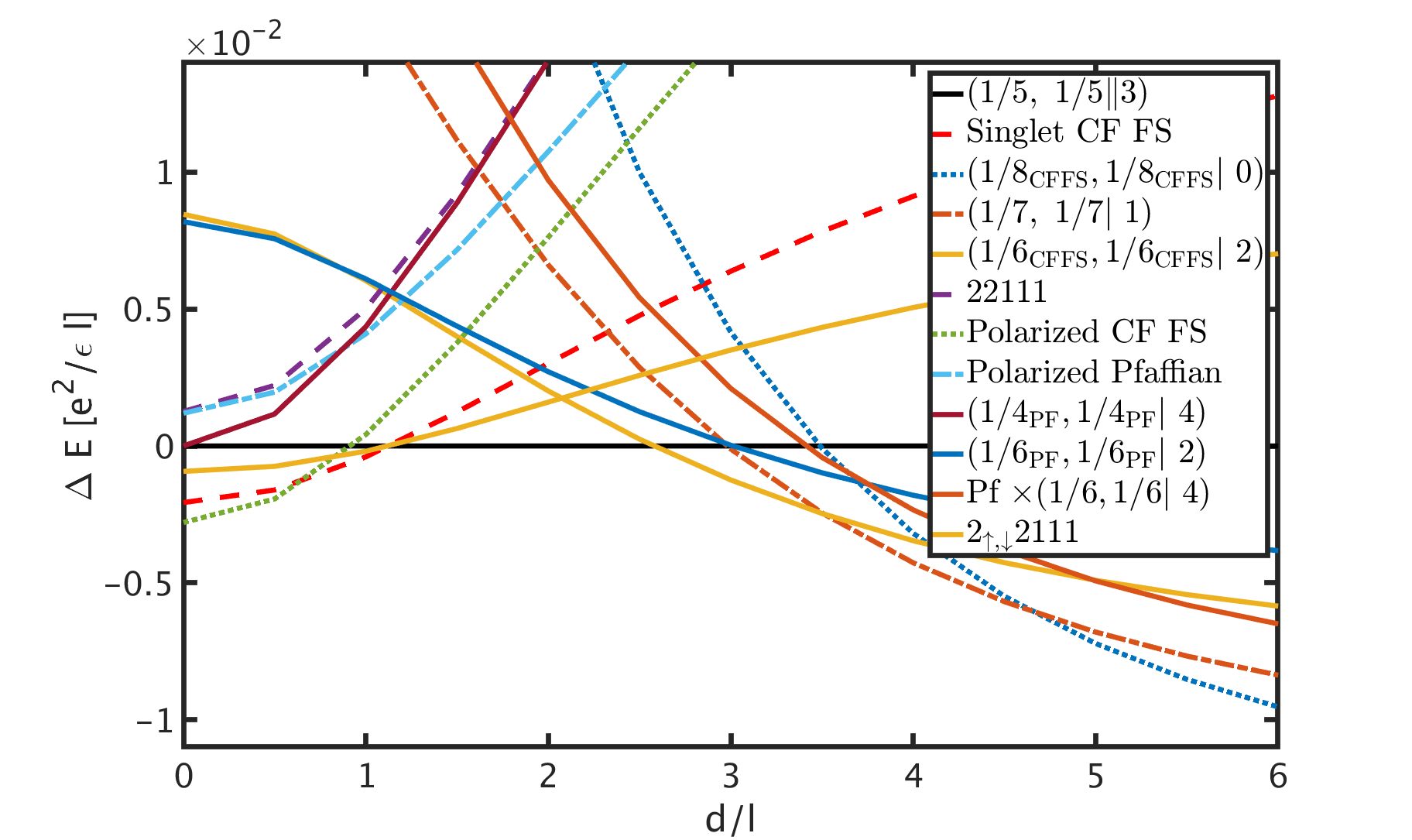}
\caption{Energy difference of bilayer liquid states from the $(1/5,~1/5|\ 3)$ state as a function of layer separation in the thermodynamic limit.}\label{fig:bliq}
\end{figure}

\begin{figure}
\includegraphics[width=3.5in]{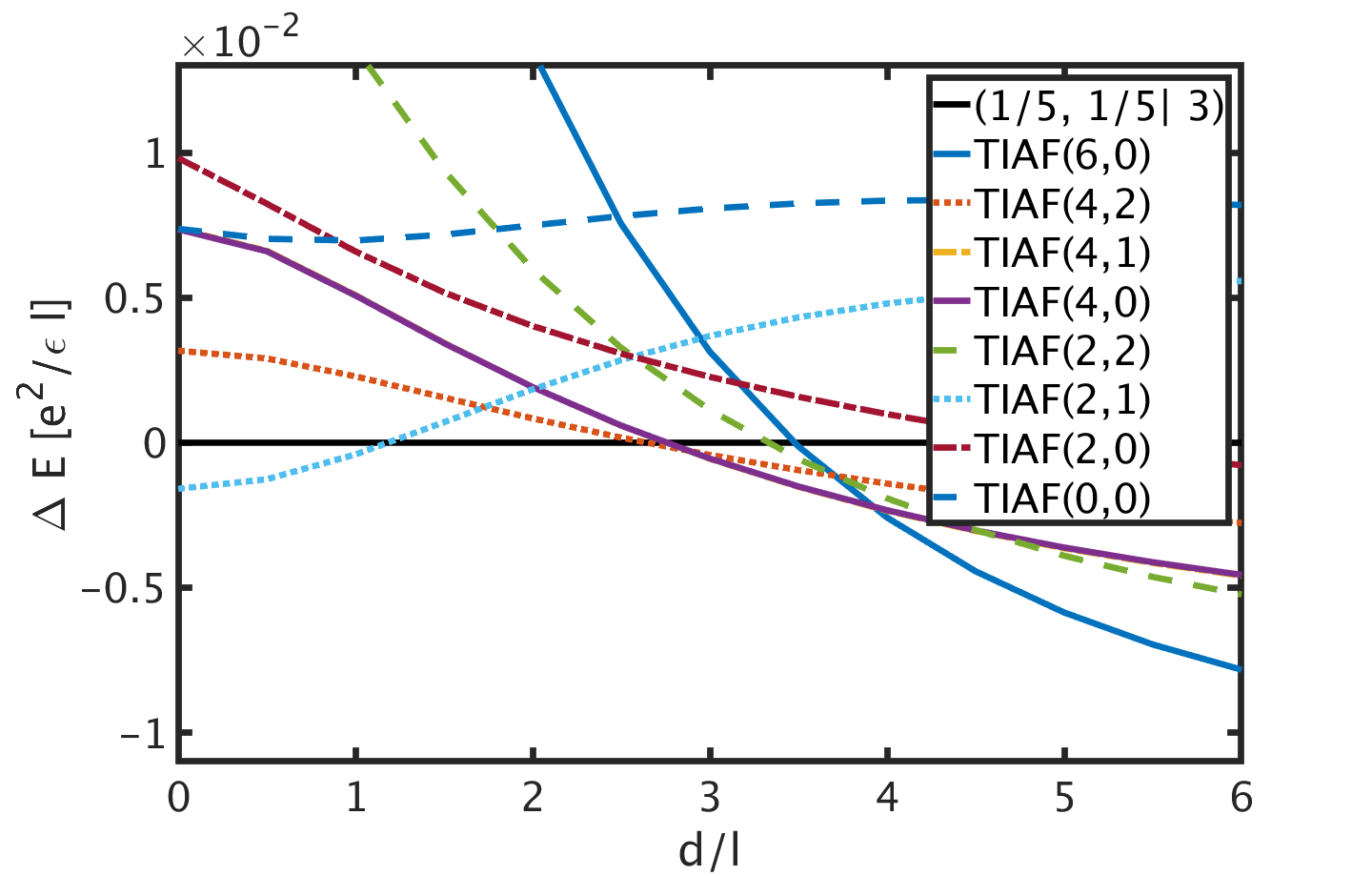}
\includegraphics[width=3.5in]{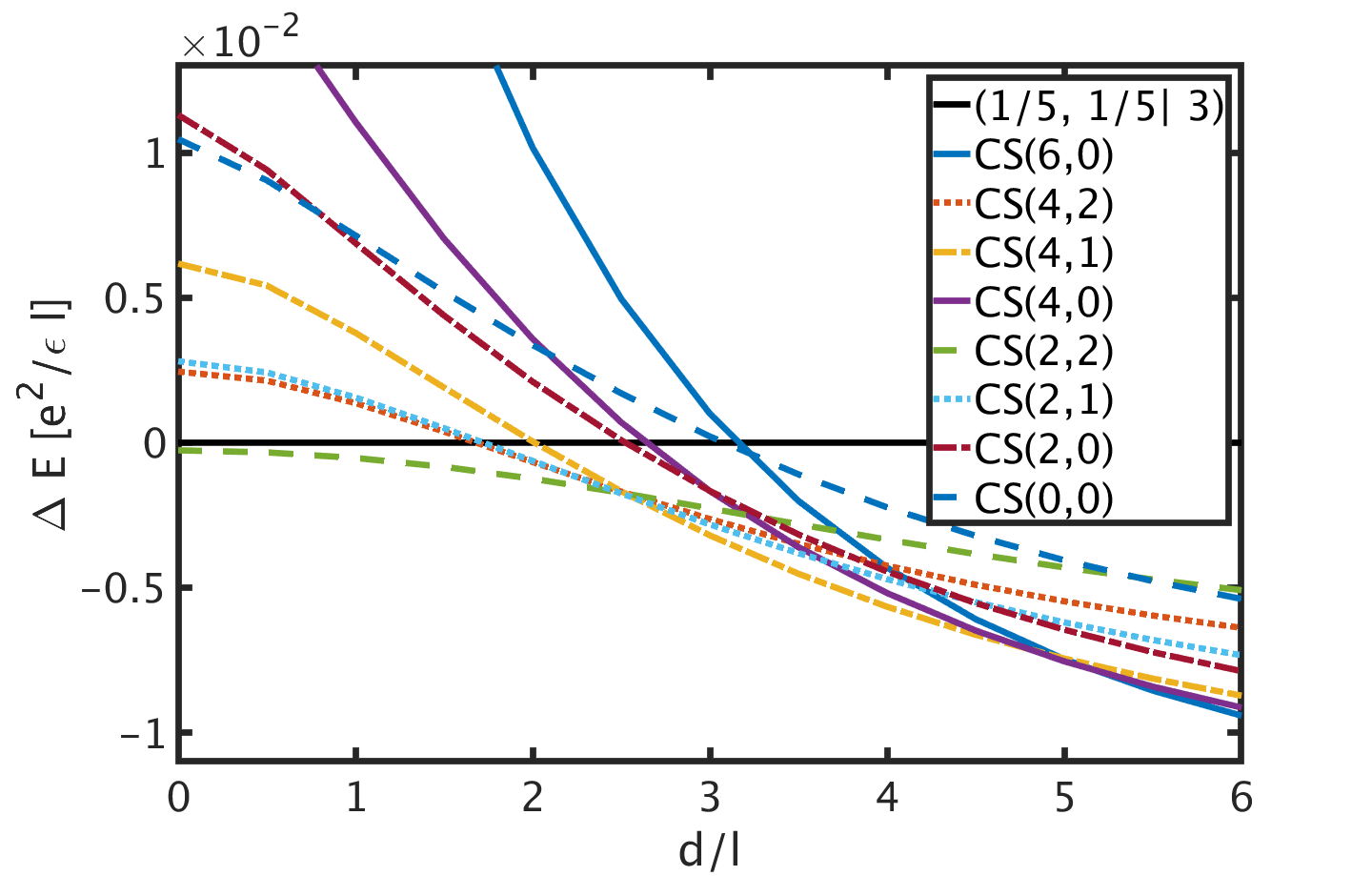}
\includegraphics[width=3.5in]{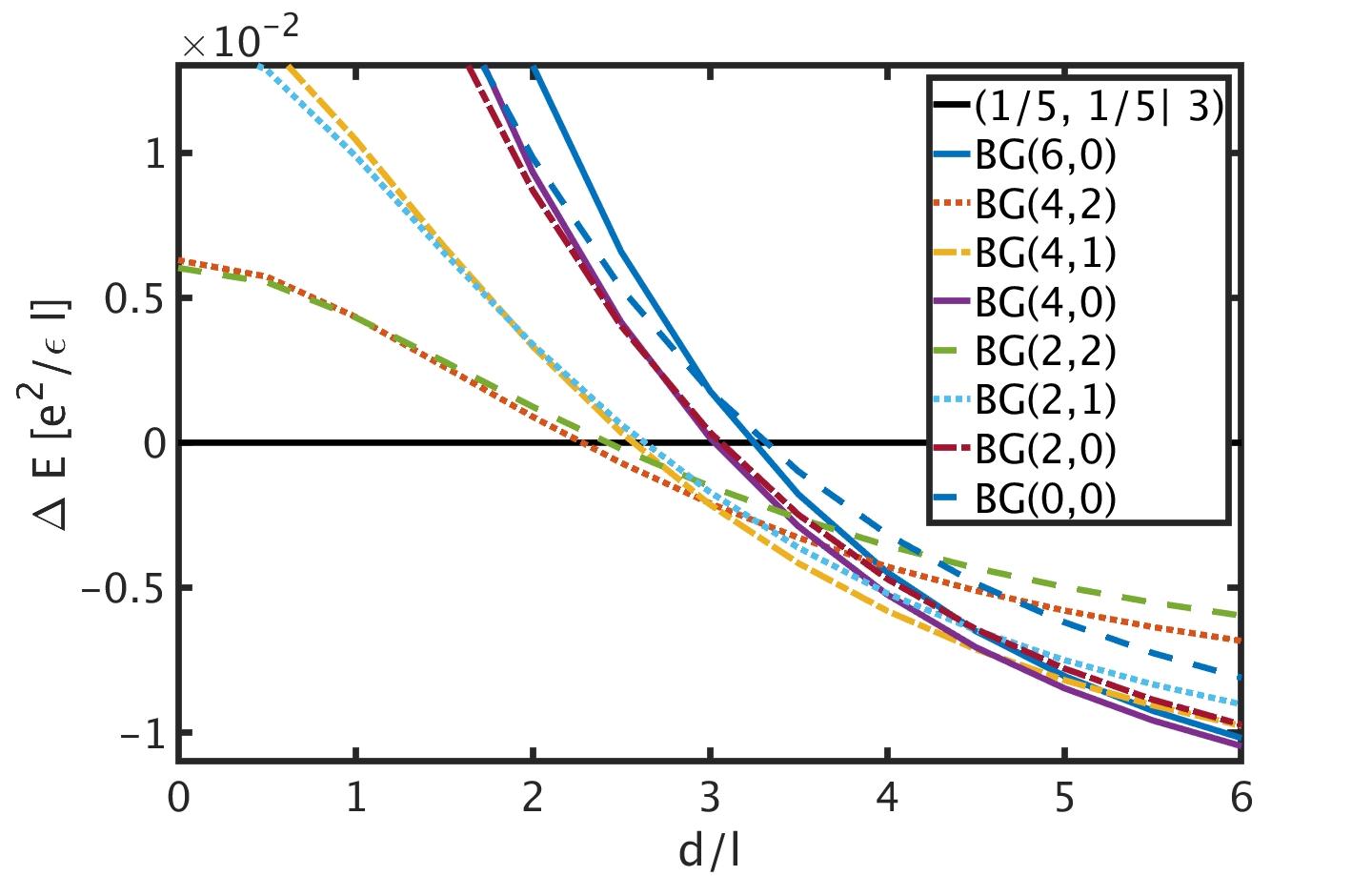}
\caption{Energy difference of TIAF (upper panel), CS (middle panel) and BG (bottom panel) crystal states from the $(1/5,~1/5|\ 3)$ state as a function of layer separation in the thermodynamic limit. The legend gives the values of $(2p,m)$ corresponding to each crystal state.}\label{fig:bilayer_crystals}
\end{figure}

The thermodynamic limit extrapolations of the energy at a separation of $d/l=0.5$ for each candidate state relative the the $(1/5,~1/5|\ 3)$ are displayed in Figs.~\ref{fig:bliqTL} and \ref{fig:bilayer_crystalsTL}. The thermodynamic energies as a function of layer separation are shown in Figs.~\ref{fig:bliq} and \ref{fig:bilayer_crystals}. The summary of the ground states can be found in lower panel of Fig.~1(c) of the main text.
%%%%%%%%%%%%%%%%%%%%

%%%%%%%%%%%%%%%%%%%%%

\section{Topological properties of the $22111$ parton state derived from its effective edge theory}
At filling fraction $\nu = 1/4$, a natural set of FQH states to consider correspond to attaching four vortices to each electron to obtain a composite fermion that sees zero magnetic field on average. Equivalently, we can consider a parton construction
\begin{align}
\wp =  b \psi,
\end{align}
where $b$ is a boson that forms a $1/4$ bosonic Laughlin FQH state, while $\psi$ is the composite fermion. If we specialize to the case where $\psi$ forms a paired state, we need to consider pairing only in odd angular momentum $\ell$ channels. This leads to wave functions of the form
\begin{align}
\Psi^{\rm CF-paired}_\ell = \Psi^{\rm paired}_\ell \Phi_1^4,
\label{eq:paired_CF_states_1_4}
\end{align}
where $\Psi^{\rm paired}_\ell$ is the wave function of an angular momentum $\ell$ paired superconductor of spinless fermions. This describes a state with central charge
\begin{align}
c = {\ell \over 2} + 1 .
  \end{align}
The edge theory consists of $\ell$ chiral Majorana modes, together with a charge mode described by a single chiral boson.

All such states have 12 topologically degenerate ground states on a torus. Correspondingly, there are $12$ topologically inequivalent quasiparticle excitations. Their charge and topological twists are summarized in Table \ref{cfprops}. For all $\ell$, the fusion rules are equivalent, so the only distinction in the topological order is the value of the topological twists and related overall Abelian phases obtained under braiding. 

We can see that the quasiparticles split up into two sectors. There are the Abelian quasiparticles, which can be obtained by inserting $2\pi a $ units of flux. These have charge $Q_a = a \nu = a/4$ and topological twist $e^{i\theta_a} = e^{i 2\pi a^2 \nu/2} =e^{i\pi a^2/4}$. We can label these particles as $V_a$. When $a = 4$, we see that we obtain a charge $1$ boson. Binding this with an electron, we get $V_a c$, which is a neutral fermion, and which corresponds to the BdG quasiparticles of the composite fermion paired state. When $a = 8$, we obtain a charge $2$ boson, which is topologically trivial.

\begin{table}
\centering
\begin{tabular}{l | r r r}
\hline
 & Charge, $Q_a$ & Twist, $e^{i \theta_a}$ \\
\hline
$V_0$  & $0$ & $1$ \\
$V_1$  & $1/4$ & $e^{i \pi/4}$ \\
$V_2$  & $1/2$ & $-1$ \\
$V_3$  & $3/4$ & $e^{i \pi/4}$ \\
$V_4$  & $1$ & $1$ \\
$V_5$  & $5/4$ &$e^{i \pi 1/4}$ \\
$V_6$  & $6/4$ & $ -1 $ \\
$V_7$  & $7/4$ & $e^{i \pi /4}$ \\
 \hline
$\sigma $  & $1/8$ & $e^{i 2\pi ( \ell/16 + 1/32)}$ \\
$\sigma V_1$  & $3/8$ &  $e^{i 2\pi ( \ell/16 + 9/32)}$\\
$\sigma V_2$  & $5/8$ &  $e^{i 2\pi ( \ell/16 + 25/32)}$\\
$\sigma V_3$  & $7/8$ & $e^{i 2\pi ( \ell/16 + 17/32)}$ \\
\hline
\end{tabular}
\caption{\label{cfprops}Twelve topologically inequivalent excitations for states of the form given in Eq.~(\ref{eq:paired_CF_states_1_4}) at $\nu = 1/4$. Binding with an electron gives rise to a topologically equivalent excitation but which differs in topological twist by $-1$. $V_4 c$ is the charge neutral composite fermion excitation. }
\end{table}

\subsection{Edge scaling dimensions and tunneling exponents}

\subsubsection{Electron operator}

The edge theory has $\ell$ chiral Majorana fermions, which we can label $\eta_i$, for $i = 1,\cdots, \ell$, and one chiral
boson $\varphi$ that carries the charge. Consequently, we have $\ell$ different types of electron operators:
\begin{align}
\Psi_{e;i} \propto \eta_i e^{i \sqrt{\nu^{-1}} \varphi}
\end{align}

More specifically, what this means is that the electron operator is represented in the low energy edge theory by a linear combination
of the above operators:
\begin{align}
\Psi_e = e^{i \sqrt{\nu^{-1}} \varphi} \sum_{i = 1}^\ell a_i \eta_i + \cdots,
\end{align}
where the $a_i$ are some constant coefficients for the expansion of the electron in terms of long wavelength field operators, and the $\cdots$
indicate higher order (less relevant) operators in the expansion.

For $\ell > 0$, this operator has scaling dimensions
\begin{align}
 (h,\bar{h}) = \left(\frac{1}{2\nu} + \frac{1}{2} ,0\right)
\end{align}
where $h$ and $\bar{h}$ are the left and right scaling dimensions. The tunneling current goes like
\begin{align}
I \propto V^{4(h+\bar{h}) - 1} = V^{9}
\end{align}
Since the edge theory is fully chiral, these exponents are quantized and universal. Importantly we see that electron tunneling does not distinguish
states with different $\ell$. 

For $\ell < 0$, this operator has scaling dimensions
\begin{align}
 (h,\bar{h}) = \left(\frac{1}{2\nu},\frac{1}{2}\right)
\end{align}
In this theory we would still get the same tunneling exponent as the case $\ell > 0$, namely
$I \propto V^{4(h+\bar{h}) - 1} = V^{9}$. However in this case the tunneling exponents are not universal and thus not quantized. In particular,
we can consider perturbations
\begin{align}
 \label{marginalDef}
\delta L = i \alpha_{ij} \partial \varphi \eta_i \partial \eta_j ,
\end{align}
for coupling constants $\alpha_{ij}$. These perturbations are marginal, having scaling dimension two, and can change the exponents of the
electron and quasiparticle operators. 
  
\subsubsection{Quasiparticle operators}

The minimally charged quasiparticle is the one carrying charge $1/8$, and is the one that is expected to dominate tunneling at a quantum point contact separating
two edges of the \it same \rm quantum Hall fluid.

For $\ell > 0$, this operator has scaling dimensions
\begin{align}
 (h,\bar{h}) = \left(\frac{1}{32} + {\ell\over 16},0\right)
\end{align}
where $h$ and $\bar{h}$ are the left and right scaling dimensions.

This implies that at a QPC, the backscattering tunneling current would be
\begin{align}
I \propto V^{4(h+\bar{h}) - 1} = V^{(2l+1)/8 - 1} = V^{(2l-7)/8}
\end{align}
For $\ell > 0$, this gives a quantized, universal value for the tunneling exponent due to charge $1/8$ quasiparticles. 

For $\ell < 0$, this operator has scaling dimensions
\begin{align}
 (h,\bar{h}) = \left(\frac{1}{32},{|\ell | \over 16} \right)
\end{align}
In the unperturbed edge theory, we would therefore expect $I \propto V^{4(h+\bar{h}) - 1} = V^{(2|\ell|+1)/8 - 1} = V^{(2|\ell|-7)/8}$. However, as in the discussion above, since the theory is not fully chiral there are marginal perturbations of the edge theory that can modify the scaling
dimensions. Therefore for $\ell < 0$ we do not expect these exponents to be universal.

We note that in the discussions of the Anti-Pfaffian state at $\nu = 1/2$ \cite{Levin07,Lee07}, it was assumed that in the presence of disorder, the coefficients
$\alpha_{ij}$ could essentially be taken to be drawn from a Gaussian distribution with zero mean. In this case such perturbations would actually
be irrelevant and a universal value of tunneling exponents would be expected. However we have no reason to consider any particular structure to the
perturbations $\alpha_{ij}$.

\subsection{Relation to other descriptions}

\subsubsection{Pfaffian}
The usual Pfaffian state has central charge $c = 1 + 1/2$. As such it corresponds to the case of $\ell = 1$ pairing of composite fermions.

\subsubsection{PH-Pfaffian}
The ``PH-Pfaffian'' state at $\nu = 1/4$ has $c = 1/2$, which corresponds to $\ell = -1$ pairing of composite fermions. 

\subsubsection{$22111$ state}
This state has central charge $c = 3/2 + 1$. As such it corresponds to angular momentum $\ell = 3$ pairing of composite
fermions (i.e. $f$-wave pairing). 

\subsubsection{$\bar{2}\bar{2}11111$}
This state has central charge $c = -3/2 + 1 = -1/2$. It is analogous to the anti-Pfaffian state, but at filling $\nu = 1/4$ instead of $\nu = 1/2$. That is, it is in the same universality class as the wave function $\Psi_{\text{APf, 1/2}} \Phi_1^2$. This corresponds to $\ell = -3$ pairing of composite fermions.

%\bibliography{../../Latex-Revtex-etc./biblio_fqhe}
%\bibliography{biblio_fqhe.bib}

\begin{thebibliography}{57}
\expandafter\ifx\csname natexlab\endcsname\relax\def\natexlab#1{#1}\fi
\expandafter\ifx\csname bibnamefont\endcsname\relax
  \def\bibnamefont#1{#1}\fi
\expandafter\ifx\csname bibfnamefont\endcsname\relax
  \def\bibfnamefont#1{#1}\fi
\expandafter\ifx\csname citenamefont\endcsname\relax
  \def\citenamefont#1{#1}\fi
\expandafter\ifx\csname url\endcsname\relax
  \def\url#1{\texttt{#1}}\fi
\expandafter\ifx\csname urlprefix\endcsname\relax\def\urlprefix{URL }\fi
\providecommand{\bibinfo}[2]{#2}
\providecommand{\eprint}[2][]{\url{#2}}

\bibitem[{\citenamefont{Moore and Read}(1991)}]{Moore91}
\bibinfo{author}{\bibfnamefont{G.}~\bibnamefont{Moore}} \bibnamefont{and}
  \bibinfo{author}{\bibfnamefont{N.}~\bibnamefont{Read}},
  \bibinfo{journal}{Nucl. Phys. B} \textbf{\bibinfo{volume}{360}},
  \bibinfo{pages}{362 } (\bibinfo{year}{1991}), ISSN \bibinfo{issn}{0550-3213},
  \urlprefix\url{http://www.sciencedirect.com/science/article/pii/055032139190%
407O}.

\bibitem[{\citenamefont{Willett et~al.}(1987)\citenamefont{Willett, Eisenstein,
  St\"ormer, Tsui, Gossard, and English}}]{Willett87}
\bibinfo{author}{\bibfnamefont{R.}~\bibnamefont{Willett}},
  \bibinfo{author}{\bibfnamefont{J.~P.} \bibnamefont{Eisenstein}},
  \bibinfo{author}{\bibfnamefont{H.~L.} \bibnamefont{St\"ormer}},
  \bibinfo{author}{\bibfnamefont{D.~C.} \bibnamefont{Tsui}},
  \bibinfo{author}{\bibfnamefont{A.~C.} \bibnamefont{Gossard}},
  \bibnamefont{and} \bibinfo{author}{\bibfnamefont{J.~H.}
  \bibnamefont{English}}, \bibinfo{journal}{Phys. Rev. Lett.}
  \textbf{\bibinfo{volume}{59}}, \bibinfo{pages}{1776} (\bibinfo{year}{1987}),
  \urlprefix\url{http://link.aps.org/doi/10.1103/PhysRevLett.59.1776}.

\bibitem[{\citenamefont{Read and Green}(2000)}]{Read00}
\bibinfo{author}{\bibfnamefont{N.}~\bibnamefont{Read}} \bibnamefont{and}
  \bibinfo{author}{\bibfnamefont{D.}~\bibnamefont{Green}},
  \bibinfo{journal}{Phys. Rev. B} \textbf{\bibinfo{volume}{61}},
  \bibinfo{pages}{10267} (\bibinfo{year}{2000}),
  \urlprefix\url{http://link.aps.org/doi/10.1103/PhysRevB.61.10267}.

\bibitem[{\citenamefont{Jain}(1989{\natexlab{a}})}]{Jain89}
\bibinfo{author}{\bibfnamefont{J.~K.} \bibnamefont{Jain}},
  \bibinfo{journal}{Phys. Rev. Lett.} \textbf{\bibinfo{volume}{63}},
  \bibinfo{pages}{199} (\bibinfo{year}{1989}{\natexlab{a}}),
  \urlprefix\url{http://link.aps.org/doi/10.1103/PhysRevLett.63.199}.

\bibitem[{\citenamefont{Jain}(2007)}]{Jain07}
\bibinfo{author}{\bibfnamefont{J.~K.} \bibnamefont{Jain}},
  \emph{\bibinfo{title}{Composite Fermions}} (\bibinfo{publisher}{Cambridge
  University Press, New York, US}, \bibinfo{year}{2007}).

\bibitem[{\citenamefont{Radu et~al.}(2008)\citenamefont{Radu, Miller, Marcus,
  Kastner, Pfeiffer, and West}}]{Radu08}
\bibinfo{author}{\bibfnamefont{I.~P.} \bibnamefont{Radu}},
  \bibinfo{author}{\bibfnamefont{J.~B.} \bibnamefont{Miller}},
  \bibinfo{author}{\bibfnamefont{C.~M.} \bibnamefont{Marcus}},
  \bibinfo{author}{\bibfnamefont{M.~A.} \bibnamefont{Kastner}},
  \bibinfo{author}{\bibfnamefont{L.~N.} \bibnamefont{Pfeiffer}},
  \bibnamefont{and} \bibinfo{author}{\bibfnamefont{K.~W.} \bibnamefont{West}},
  \bibinfo{journal}{SCIENCE}  (\bibinfo{year}{2008}).

\bibitem[{\citenamefont{Willett et~al.}(2009)\citenamefont{Willett, Pfeiffer,
  and West}}]{Willett09}
\bibinfo{author}{\bibfnamefont{R.~L.} \bibnamefont{Willett}},
  \bibinfo{author}{\bibfnamefont{L.~N.} \bibnamefont{Pfeiffer}},
  \bibnamefont{and} \bibinfo{author}{\bibfnamefont{K.~W.} \bibnamefont{West}},
  \bibinfo{journal}{Proceedings of the National Academy of Sciences}
  \textbf{\bibinfo{volume}{106}}, \bibinfo{pages}{8853} (\bibinfo{year}{2009}),
  \eprint{http://www.pnas.org/content/106/22/8853.full.pdf},
  \urlprefix\url{http://www.pnas.org/content/106/22/8853.abstract}.

\bibitem[{\citenamefont{Willett et~al.}(2010)\citenamefont{Willett, Pfeiffer,
  and West}}]{Willett10}
\bibinfo{author}{\bibfnamefont{R.~L.} \bibnamefont{Willett}},
  \bibinfo{author}{\bibfnamefont{L.~N.} \bibnamefont{Pfeiffer}},
  \bibnamefont{and} \bibinfo{author}{\bibfnamefont{K.~W.} \bibnamefont{West}},
  \bibinfo{journal}{Phys. Rev. B} \textbf{\bibinfo{volume}{82}},
  \bibinfo{pages}{205301} (\bibinfo{year}{2010}),
  \urlprefix\url{http://link.aps.org/doi/10.1103/PhysRevB.82.205301}.

\bibitem[{\citenamefont{Banerjee et~al.}(2017)\citenamefont{Banerjee, Heiblum,
  Rosenblatt, Oreg, Feldman, Stern, and Umansky}}]{Banerjee17}
\bibinfo{author}{\bibfnamefont{M.}~\bibnamefont{Banerjee}},
  \bibinfo{author}{\bibfnamefont{M.}~\bibnamefont{Heiblum}},
  \bibinfo{author}{\bibfnamefont{A.}~\bibnamefont{Rosenblatt}},
  \bibinfo{author}{\bibfnamefont{Y.}~\bibnamefont{Oreg}},
  \bibinfo{author}{\bibfnamefont{D.~E.} \bibnamefont{Feldman}},
  \bibinfo{author}{\bibfnamefont{A.}~\bibnamefont{Stern}}, \bibnamefont{and}
  \bibinfo{author}{\bibfnamefont{V.}~\bibnamefont{Umansky}},
  \bibinfo{journal}{Nature} \textbf{\bibinfo{volume}{545}},
  \bibinfo{pages}{75+} (\bibinfo{year}{2017}), ISSN \bibinfo{issn}{0028-0836}.

\bibitem[{\citenamefont{Banerjee et~al.}(2018)\citenamefont{Banerjee, Heiblum,
  Umansky, Feldman, Oreg, and Stern}}]{Banerjee17b}
\bibinfo{author}{\bibfnamefont{M.}~\bibnamefont{Banerjee}},
  \bibinfo{author}{\bibfnamefont{M.}~\bibnamefont{Heiblum}},
  \bibinfo{author}{\bibfnamefont{V.}~\bibnamefont{Umansky}},
  \bibinfo{author}{\bibfnamefont{D.~E.} \bibnamefont{Feldman}},
  \bibinfo{author}{\bibfnamefont{Y.}~\bibnamefont{Oreg}}, \bibnamefont{and}
  \bibinfo{author}{\bibfnamefont{A.}~\bibnamefont{Stern}},
  \bibinfo{journal}{Nature} \textbf{\bibinfo{volume}{559}},
  \bibinfo{pages}{205} (\bibinfo{year}{2018}), ISSN \bibinfo{issn}{1476-4687},
  \urlprefix\url{https://doi.org/10.1038/s41586-018-0184-1}.

\bibitem[{\citenamefont{Levin et~al.}(2007)\citenamefont{Levin, Halperin, and
  Rosenow}}]{Levin07}
\bibinfo{author}{\bibfnamefont{M.}~\bibnamefont{Levin}},
  \bibinfo{author}{\bibfnamefont{B.~I.} \bibnamefont{Halperin}},
  \bibnamefont{and} \bibinfo{author}{\bibfnamefont{B.}~\bibnamefont{Rosenow}},
  \bibinfo{journal}{Phys. Rev. Lett.} \textbf{\bibinfo{volume}{99}},
  \bibinfo{pages}{236806} (\bibinfo{year}{2007}),
  \urlprefix\url{http://link.aps.org/doi/10.1103/PhysRevLett.99.236806}.

\bibitem[{\citenamefont{Lee et~al.}(2007)\citenamefont{Lee, Ryu, Nayak, and
  Fisher}}]{Lee07}
\bibinfo{author}{\bibfnamefont{S.-S.} \bibnamefont{Lee}},
  \bibinfo{author}{\bibfnamefont{S.}~\bibnamefont{Ryu}},
  \bibinfo{author}{\bibfnamefont{C.}~\bibnamefont{Nayak}}, \bibnamefont{and}
  \bibinfo{author}{\bibfnamefont{M.~P.~A.} \bibnamefont{Fisher}},
  \bibinfo{journal}{Phys. Rev. Lett.} \textbf{\bibinfo{volume}{99}},
  \bibinfo{pages}{236807} (\bibinfo{year}{2007}),
  \urlprefix\url{http://link.aps.org/doi/10.1103/PhysRevLett.99.236807}.

\bibitem[{\citenamefont{Luhman et~al.}(2008)\citenamefont{Luhman, Pan, Tsui,
  Pfeiffer, Baldwin, and West}}]{Luhman08}
\bibinfo{author}{\bibfnamefont{D.~R.} \bibnamefont{Luhman}},
  \bibinfo{author}{\bibfnamefont{W.}~\bibnamefont{Pan}},
  \bibinfo{author}{\bibfnamefont{D.~C.} \bibnamefont{Tsui}},
  \bibinfo{author}{\bibfnamefont{L.~N.} \bibnamefont{Pfeiffer}},
  \bibinfo{author}{\bibfnamefont{K.~W.} \bibnamefont{Baldwin}},
  \bibnamefont{and} \bibinfo{author}{\bibfnamefont{K.~W.} \bibnamefont{West}},
  \bibinfo{journal}{Phys. Rev. Lett.} \textbf{\bibinfo{volume}{101}},
  \bibinfo{pages}{266804} (\bibinfo{year}{2008}),
  \urlprefix\url{https://link.aps.org/doi/10.1103/PhysRevLett.101.266804}.

\bibitem[{\citenamefont{Shabani
  et~al.}(2009{\natexlab{a}})\citenamefont{Shabani, Gokmen, and
  Shayegan}}]{Shabani09a}
\bibinfo{author}{\bibfnamefont{J.}~\bibnamefont{Shabani}},
  \bibinfo{author}{\bibfnamefont{T.}~\bibnamefont{Gokmen}}, \bibnamefont{and}
  \bibinfo{author}{\bibfnamefont{M.}~\bibnamefont{Shayegan}},
  \bibinfo{journal}{Phys. Rev. Lett.} \textbf{\bibinfo{volume}{103}},
  \bibinfo{pages}{046805} (\bibinfo{year}{2009}{\natexlab{a}}),
  \urlprefix\url{https://link.aps.org/doi/10.1103/PhysRevLett.103.046805}.

\bibitem[{\citenamefont{Shabani
  et~al.}(2009{\natexlab{b}})\citenamefont{Shabani, Gokmen, Chiu, and
  Shayegan}}]{Shabani09b}
\bibinfo{author}{\bibfnamefont{J.}~\bibnamefont{Shabani}},
  \bibinfo{author}{\bibfnamefont{T.}~\bibnamefont{Gokmen}},
  \bibinfo{author}{\bibfnamefont{Y.~T.} \bibnamefont{Chiu}}, \bibnamefont{and}
  \bibinfo{author}{\bibfnamefont{M.}~\bibnamefont{Shayegan}},
  \bibinfo{journal}{Phys. Rev. Lett.} \textbf{\bibinfo{volume}{103}},
  \bibinfo{pages}{256802} (\bibinfo{year}{2009}{\natexlab{b}}),
  \urlprefix\url{https://link.aps.org/doi/10.1103/PhysRevLett.103.256802}.

\bibitem[{\citenamefont{Shabani et~al.}(2013)\citenamefont{Shabani, Liu,
  Shayegan, Pfeiffer, West, and Baldwin}}]{Shabani13}
\bibinfo{author}{\bibfnamefont{J.}~\bibnamefont{Shabani}},
  \bibinfo{author}{\bibfnamefont{Y.}~\bibnamefont{Liu}},
  \bibinfo{author}{\bibfnamefont{M.}~\bibnamefont{Shayegan}},
  \bibinfo{author}{\bibfnamefont{L.~N.} \bibnamefont{Pfeiffer}},
  \bibinfo{author}{\bibfnamefont{K.~W.} \bibnamefont{West}}, \bibnamefont{and}
  \bibinfo{author}{\bibfnamefont{K.~W.} \bibnamefont{Baldwin}},
  \bibinfo{journal}{Phys. Rev. B} \textbf{\bibinfo{volume}{88}},
  \bibinfo{pages}{245413} (\bibinfo{year}{2013}),
  \urlprefix\url{https://link.aps.org/doi/10.1103/PhysRevB.88.245413}.

\bibitem[{\citenamefont{Jain}(1989{\natexlab{b}})}]{Jain89b}
\bibinfo{author}{\bibfnamefont{J.~K.} \bibnamefont{Jain}},
  \bibinfo{journal}{Phys. Rev. B} \textbf{\bibinfo{volume}{40}},
  \bibinfo{pages}{8079} (\bibinfo{year}{1989}{\natexlab{b}}),
  \urlprefix\url{http://link.aps.org/doi/10.1103/PhysRevB.40.8079}.

\bibitem[{\citenamefont{Jain}(1990)}]{Jain90}
\bibinfo{author}{\bibfnamefont{J.~K.} \bibnamefont{Jain}},
  \bibinfo{journal}{Phys. Rev. B} \textbf{\bibinfo{volume}{41}},
  \bibinfo{pages}{7653} (\bibinfo{year}{1990}).

\bibitem[{\citenamefont{Blok and Wen}(1990{\natexlab{a}})}]{Blok90}
\bibinfo{author}{\bibfnamefont{B.}~\bibnamefont{Blok}} \bibnamefont{and}
  \bibinfo{author}{\bibfnamefont{X.~G.} \bibnamefont{Wen}},
  \bibinfo{journal}{Phys. Rev. B} \textbf{\bibinfo{volume}{42}},
  \bibinfo{pages}{8145} (\bibinfo{year}{1990}{\natexlab{a}}),
  \urlprefix\url{http://link.aps.org/doi/10.1103/PhysRevB.42.8145}.

\bibitem[{\citenamefont{Blok and Wen}(1990{\natexlab{b}})}]{Blok90b}
\bibinfo{author}{\bibfnamefont{B.}~\bibnamefont{Blok}} \bibnamefont{and}
  \bibinfo{author}{\bibfnamefont{X.~G.} \bibnamefont{Wen}},
  \bibinfo{journal}{Phys. Rev. B} \textbf{\bibinfo{volume}{42}},
  \bibinfo{pages}{8133} (\bibinfo{year}{1990}{\natexlab{b}}),
  \urlprefix\url{http://link.aps.org/doi/10.1103/PhysRevB.42.8133}.

\bibitem[{\citenamefont{Wen}(1991)}]{Wen91}
\bibinfo{author}{\bibfnamefont{X.~G.} \bibnamefont{Wen}},
  \bibinfo{journal}{Phys. Rev. Lett.} \textbf{\bibinfo{volume}{66}},
  \bibinfo{pages}{802} (\bibinfo{year}{1991}),
  \urlprefix\url{http://link.aps.org/doi/10.1103/PhysRevLett.66.802}.

\bibitem[{\citenamefont{Wen}(1992)}]{Wen92b}
\bibinfo{author}{\bibfnamefont{X.-G.} \bibnamefont{Wen}},
  \bibinfo{journal}{International Journal of Modern Physics B}
  \textbf{\bibinfo{volume}{06}}, \bibinfo{pages}{1711} (\bibinfo{year}{1992}),
  \urlprefix\url{http://www.worldscientific.com/doi/abs/10.1142/S0217979292000%
840}.

\bibitem[{\citenamefont{Wu et~al.}(2017)\citenamefont{Wu, Shi, and
  Jain}}]{Wu16}
\bibinfo{author}{\bibfnamefont{Y.}~\bibnamefont{Wu}},
  \bibinfo{author}{\bibfnamefont{T.}~\bibnamefont{Shi}}, \bibnamefont{and}
  \bibinfo{author}{\bibfnamefont{J.~K.} \bibnamefont{Jain}},
  \bibinfo{journal}{Nano Letters} \textbf{\bibinfo{volume}{17}},
  \bibinfo{pages}{4643} (\bibinfo{year}{2017}), \bibinfo{note}{pMID: 28649831},
  \eprint{http://dx.doi.org/10.1021/acs.nanolett.7b01080},
  \urlprefix\url{http://dx.doi.org/10.1021/acs.nanolett.7b01080}.

\bibitem[{\citenamefont{Bandyopadhyay et~al.}(2018)\citenamefont{Bandyopadhyay,
  Chen, Ahari, Ortiz, Nussinov, and Seidel}}]{Bandyopadhyay18}
\bibinfo{author}{\bibfnamefont{S.}~\bibnamefont{Bandyopadhyay}},
  \bibinfo{author}{\bibfnamefont{L.}~\bibnamefont{Chen}},
  \bibinfo{author}{\bibfnamefont{M.~T.} \bibnamefont{Ahari}},
  \bibinfo{author}{\bibfnamefont{G.}~\bibnamefont{Ortiz}},
  \bibinfo{author}{\bibfnamefont{Z.}~\bibnamefont{Nussinov}}, \bibnamefont{and}
  \bibinfo{author}{\bibfnamefont{A.}~\bibnamefont{Seidel}},
  \bibinfo{journal}{Phys. Rev. B} \textbf{\bibinfo{volume}{98}},
  \bibinfo{pages}{161118} (\bibinfo{year}{2018}),
  \urlprefix\url{https://link.aps.org/doi/10.1103/PhysRevB.98.161118}.

\bibitem[{\citenamefont{W\'ojs}(2009)}]{Wojs09}
\bibinfo{author}{\bibfnamefont{A.}~\bibnamefont{W\'ojs}},
  \bibinfo{journal}{Phys. Rev. B} \textbf{\bibinfo{volume}{80}},
  \bibinfo{pages}{041104} (\bibinfo{year}{2009}),
  \urlprefix\url{http://link.aps.org/doi/10.1103/PhysRevB.80.041104}.

\bibitem[{\citenamefont{Balram et~al.}(2018)\citenamefont{Balram, Barkeshli,
  and Rudner}}]{Balram18}
\bibinfo{author}{\bibfnamefont{A.~C.} \bibnamefont{Balram}},
  \bibinfo{author}{\bibfnamefont{M.}~\bibnamefont{Barkeshli}},
  \bibnamefont{and} \bibinfo{author}{\bibfnamefont{M.~S.}
  \bibnamefont{Rudner}}, \bibinfo{journal}{Phys. Rev. B}
  \textbf{\bibinfo{volume}{98}}, \bibinfo{pages}{035127}
  (\bibinfo{year}{2018}),
  \urlprefix\url{https://link.aps.org/doi/10.1103/PhysRevB.98.035127}.

\bibitem[{\citenamefont{Kim et~al.}(2019)\citenamefont{Kim, Balram, Taniguchi,
  Watanabe, Jain, and Smet}}]{Kim18}
\bibinfo{author}{\bibfnamefont{Y.}~\bibnamefont{Kim}},
  \bibinfo{author}{\bibfnamefont{A.~C.} \bibnamefont{Balram}},
  \bibinfo{author}{\bibfnamefont{T.}~\bibnamefont{Taniguchi}},
  \bibinfo{author}{\bibfnamefont{K.}~\bibnamefont{Watanabe}},
  \bibinfo{author}{\bibfnamefont{J.~K.} \bibnamefont{Jain}}, \bibnamefont{and}
  \bibinfo{author}{\bibfnamefont{J.~H.} \bibnamefont{Smet}},
  \bibinfo{journal}{Nature Physics} \textbf{\bibinfo{volume}{15}},
  \bibinfo{pages}{154} (\bibinfo{year}{2019}), ISSN \bibinfo{issn}{1745-2481},
  \urlprefix\url{https://doi.org/10.1038/s41567-018-0355-x}.

\bibitem[{\citenamefont{Halperin et~al.}(1993)\citenamefont{Halperin, Lee, and
  Read}}]{Halperin93}
\bibinfo{author}{\bibfnamefont{B.~I.} \bibnamefont{Halperin}},
  \bibinfo{author}{\bibfnamefont{P.~A.} \bibnamefont{Lee}}, \bibnamefont{and}
  \bibinfo{author}{\bibfnamefont{N.}~\bibnamefont{Read}},
  \bibinfo{journal}{Phys. Rev. B} \textbf{\bibinfo{volume}{47}},
  \bibinfo{pages}{7312} (\bibinfo{year}{1993}),
  \urlprefix\url{http://link.aps.org/doi/10.1103/PhysRevB.47.7312}.

\bibitem[{\citenamefont{Rezayi and Read}(1994)}]{Rezayi94}
\bibinfo{author}{\bibfnamefont{E.}~\bibnamefont{Rezayi}} \bibnamefont{and}
  \bibinfo{author}{\bibfnamefont{N.}~\bibnamefont{Read}},
  \bibinfo{journal}{Phys. Rev. Lett.} \textbf{\bibinfo{volume}{72}},
  \bibinfo{pages}{900} (\bibinfo{year}{1994}),
  \urlprefix\url{http://link.aps.org/doi/10.1103/PhysRevLett.72.900}.

\bibitem[{\citenamefont{Jain and Kamilla}(1997{\natexlab{a}})}]{Jain97}
\bibinfo{author}{\bibfnamefont{J.~K.} \bibnamefont{Jain}} \bibnamefont{and}
  \bibinfo{author}{\bibfnamefont{R.~K.} \bibnamefont{Kamilla}},
  \bibinfo{journal}{Int. J. Mod. Phys. B} \textbf{\bibinfo{volume}{11}},
  \bibinfo{pages}{2621} (\bibinfo{year}{1997}{\natexlab{a}}).

\bibitem[{\citenamefont{Jain and Kamilla}(1997{\natexlab{b}})}]{Jain97b}
\bibinfo{author}{\bibfnamefont{J.~K.} \bibnamefont{Jain}} \bibnamefont{and}
  \bibinfo{author}{\bibfnamefont{R.~K.} \bibnamefont{Kamilla}},
  \bibinfo{journal}{Phys. Rev. B} \textbf{\bibinfo{volume}{55}},
  \bibinfo{pages}{R4895} (\bibinfo{year}{1997}{\natexlab{b}}),
  \urlprefix\url{http://link.aps.org/doi/10.1103/PhysRevB.55.R4895}.

\bibitem[{\citenamefont{Haldane}(1983)}]{Haldane83}
\bibinfo{author}{\bibfnamefont{F.~D.~M.} \bibnamefont{Haldane}},
  \bibinfo{journal}{Phys. Rev. Lett.} \textbf{\bibinfo{volume}{51}},
  \bibinfo{pages}{605} (\bibinfo{year}{1983}),
  \urlprefix\url{http://link.aps.org/doi/10.1103/PhysRevLett.51.605}.

\bibitem[{SM-()}]{SM-Faugno-14}
\bibinfo{note}{See Supplemental Material accompanying this paper, which
  includes details of numerical calculations; a more complete account of
  bilayer states; and a discussion of the quasiparticle operators.}

\bibitem[{\citenamefont{Son}(2015)}]{Son15}
\bibinfo{author}{\bibfnamefont{D.~T.} \bibnamefont{Son}},
  \bibinfo{journal}{Phys. Rev. X} \textbf{\bibinfo{volume}{5}},
  \bibinfo{pages}{031027} (\bibinfo{year}{2015}),
  \urlprefix\url{http://link.aps.org/doi/10.1103/PhysRevX.5.031027}.

\bibitem[{\citenamefont{Jolicoeur}(2007)}]{Jolicoeur07}
\bibinfo{author}{\bibfnamefont{T.}~\bibnamefont{Jolicoeur}},
  \bibinfo{journal}{Phys. Rev. Lett.} \textbf{\bibinfo{volume}{99}},
  \bibinfo{pages}{036805} (\bibinfo{year}{2007}),
  \urlprefix\url{http://link.aps.org/doi/10.1103/PhysRevLett.99.036805}.

\bibitem[{\citenamefont{Zucker and Feldman}(2016)}]{Zucker16}
\bibinfo{author}{\bibfnamefont{P.~T.} \bibnamefont{Zucker}} \bibnamefont{and}
  \bibinfo{author}{\bibfnamefont{D.~E.} \bibnamefont{Feldman}},
  \bibinfo{journal}{Phys. Rev. Lett.} \textbf{\bibinfo{volume}{117}},
  \bibinfo{pages}{096802} (\bibinfo{year}{2016}).

\bibitem[{\citenamefont{Mishmash et~al.}(2018)\citenamefont{Mishmash, Mross,
  Alicea, and Motrunich}}]{Mishmash18}
\bibinfo{author}{\bibfnamefont{R.~V.} \bibnamefont{Mishmash}},
  \bibinfo{author}{\bibfnamefont{D.~F.} \bibnamefont{Mross}},
  \bibinfo{author}{\bibfnamefont{J.}~\bibnamefont{Alicea}}, \bibnamefont{and}
  \bibinfo{author}{\bibfnamefont{O.~I.} \bibnamefont{Motrunich}},
  \bibinfo{journal}{Phys. Rev. B} \textbf{\bibinfo{volume}{98}},
  \bibinfo{pages}{081107} (\bibinfo{year}{2018}),
  \urlprefix\url{https://link.aps.org/doi/10.1103/PhysRevB.98.081107}.

\bibitem[{\citenamefont{Rowell et~al.}(2009)\citenamefont{Rowell, Stong, and
  Wang}}]{Rowell09}
\bibinfo{author}{\bibfnamefont{E.}~\bibnamefont{Rowell}},
  \bibinfo{author}{\bibfnamefont{R.}~\bibnamefont{Stong}}, \bibnamefont{and}
  \bibinfo{author}{\bibfnamefont{Z.}~\bibnamefont{Wang}},
  \bibinfo{journal}{Communications in Mathematical Physics}
  \textbf{\bibinfo{volume}{292}}, \bibinfo{pages}{343} (\bibinfo{year}{2009}),
  ISSN \bibinfo{issn}{1432-0916}.

\bibitem[{\citenamefont{Ortalano et~al.}(1997)\citenamefont{Ortalano, He, and
  Das~Sarma}}]{Ortalano97}
\bibinfo{author}{\bibfnamefont{M.~W.} \bibnamefont{Ortalano}},
  \bibinfo{author}{\bibfnamefont{S.}~\bibnamefont{He}}, \bibnamefont{and}
  \bibinfo{author}{\bibfnamefont{S.}~\bibnamefont{Das~Sarma}},
  \bibinfo{journal}{Phys. Rev. B} \textbf{\bibinfo{volume}{55}},
  \bibinfo{pages}{7702} (\bibinfo{year}{1997}),
  \urlprefix\url{http://link.aps.org/doi/10.1103/PhysRevB.55.7702}.

\bibitem[{\citenamefont{Papi\ifmmode~\acute{c}\else \'{c}\fi{}
  et~al.}(2009)\citenamefont{Papi\ifmmode~\acute{c}\else \'{c}\fi{}, M\"oller,
  Milovanovi\ifmmode~\acute{c}\else \'{c}\fi{}, Regnault, and
  Goerbig}}]{Papic09}
\bibinfo{author}{\bibfnamefont{Z.}~\bibnamefont{Papi\ifmmode~\acute{c}\else
  \'{c}\fi{}}}, \bibinfo{author}{\bibfnamefont{G.}~\bibnamefont{M\"oller}},
  \bibinfo{author}{\bibfnamefont{M.~V.}
  \bibnamefont{Milovanovi\ifmmode~\acute{c}\else \'{c}\fi{}}},
  \bibinfo{author}{\bibfnamefont{N.}~\bibnamefont{Regnault}}, \bibnamefont{and}
  \bibinfo{author}{\bibfnamefont{M.~O.} \bibnamefont{Goerbig}},
  \bibinfo{journal}{Phys. Rev. B} \textbf{\bibinfo{volume}{79}},
  \bibinfo{pages}{245325} (\bibinfo{year}{2009}),
  \urlprefix\url{https://link.aps.org/doi/10.1103/PhysRevB.79.245325}.

\bibitem[{\citenamefont{Liu et~al.}(2014)\citenamefont{Liu, Hasdemir, W\'ojs,
  Jain, Pfeiffer, West, Baldwin, and Shayegan}}]{Liu14}
\bibinfo{author}{\bibfnamefont{Y.}~\bibnamefont{Liu}},
  \bibinfo{author}{\bibfnamefont{S.}~\bibnamefont{Hasdemir}},
  \bibinfo{author}{\bibfnamefont{A.}~\bibnamefont{W\'ojs}},
  \bibinfo{author}{\bibfnamefont{J.~K.} \bibnamefont{Jain}},
  \bibinfo{author}{\bibfnamefont{L.~N.} \bibnamefont{Pfeiffer}},
  \bibinfo{author}{\bibfnamefont{K.~W.} \bibnamefont{West}},
  \bibinfo{author}{\bibfnamefont{K.~W.} \bibnamefont{Baldwin}},
  \bibnamefont{and} \bibinfo{author}{\bibfnamefont{M.}~\bibnamefont{Shayegan}},
  \bibinfo{journal}{Phys. Rev. B} \textbf{\bibinfo{volume}{90}},
  \bibinfo{pages}{085301} (\bibinfo{year}{2014}),
  \urlprefix\url{http://link.aps.org/doi/10.1103/PhysRevB.90.085301}.

\bibitem[{\citenamefont{Zhang et~al.}(2016)\citenamefont{Zhang, W\'ojs, and
  Jain}}]{Zhang16}
\bibinfo{author}{\bibfnamefont{Y.}~\bibnamefont{Zhang}},
  \bibinfo{author}{\bibfnamefont{A.}~\bibnamefont{W\'ojs}}, \bibnamefont{and}
  \bibinfo{author}{\bibfnamefont{J.~K.} \bibnamefont{Jain}},
  \bibinfo{journal}{Phys. Rev. Lett.} \textbf{\bibinfo{volume}{117}},
  \bibinfo{pages}{116803} (\bibinfo{year}{2016}),
  \urlprefix\url{http://link.aps.org/doi/10.1103/PhysRevLett.117.116803}.

\bibitem[{\citenamefont{Park and Jain}(1999)}]{Park99}
\bibinfo{author}{\bibfnamefont{K.}~\bibnamefont{Park}} \bibnamefont{and}
  \bibinfo{author}{\bibfnamefont{J.~K.} \bibnamefont{Jain}},
  \bibinfo{journal}{Phys. Rev. Lett.} \textbf{\bibinfo{volume}{83}},
  \bibinfo{pages}{5543} (\bibinfo{year}{1999}),
  \urlprefix\url{http://link.aps.org/doi/10.1103/PhysRevLett.83.5543}.

\bibitem[{\citenamefont{Balram et~al.}(2015)\citenamefont{Balram, T\"oke,
  W\'ojs, and Jain}}]{Balram15a}
\bibinfo{author}{\bibfnamefont{A.~C.} \bibnamefont{Balram}},
  \bibinfo{author}{\bibfnamefont{C.}~\bibnamefont{T\"oke}},
  \bibinfo{author}{\bibfnamefont{A.}~\bibnamefont{W\'ojs}}, \bibnamefont{and}
  \bibinfo{author}{\bibfnamefont{J.~K.} \bibnamefont{Jain}},
  \bibinfo{journal}{Phys. Rev. B} \textbf{\bibinfo{volume}{92}},
  \bibinfo{pages}{075410} (\bibinfo{year}{2015}),
  \urlprefix\url{http://link.aps.org/doi/10.1103/PhysRevB.92.075410}.

\bibitem[{\citenamefont{Faugno et~al.}(2018)\citenamefont{Faugno, Duthie,
  Wales, and Jain}}]{Faugno18}
\bibinfo{author}{\bibfnamefont{W.~N.} \bibnamefont{Faugno}},
  \bibinfo{author}{\bibfnamefont{A.~J.} \bibnamefont{Duthie}},
  \bibinfo{author}{\bibfnamefont{D.~J.} \bibnamefont{Wales}}, \bibnamefont{and}
  \bibinfo{author}{\bibfnamefont{J.~K.} \bibnamefont{Jain}},
  \bibinfo{journal}{Phys. Rev. B} \textbf{\bibinfo{volume}{97}},
  \bibinfo{pages}{245424} (\bibinfo{year}{2018}),
  \urlprefix\url{https://link.aps.org/doi/10.1103/PhysRevB.97.245424}.

\bibitem[{\citenamefont{Eisenstein et~al.}(1992)\citenamefont{Eisenstein,
  Boebinger, Pfeiffer, West, and He}}]{Eisenstein92}
\bibinfo{author}{\bibfnamefont{J.~P.} \bibnamefont{Eisenstein}},
  \bibinfo{author}{\bibfnamefont{G.~S.} \bibnamefont{Boebinger}},
  \bibinfo{author}{\bibfnamefont{L.~N.} \bibnamefont{Pfeiffer}},
  \bibinfo{author}{\bibfnamefont{K.~W.} \bibnamefont{West}}, \bibnamefont{and}
  \bibinfo{author}{\bibfnamefont{S.}~\bibnamefont{He}}, \bibinfo{journal}{Phys.
  Rev. Lett.} \textbf{\bibinfo{volume}{68}}, \bibinfo{pages}{1383}
  (\bibinfo{year}{1992}),
  \urlprefix\url{http://link.aps.org/doi/10.1103/PhysRevLett.68.1383}.

\bibitem[{\citenamefont{Liu et~al.}(2018)\citenamefont{Liu, Hao, Watanabe,
  Taniguchi, Halperin, and Kim}}]{Kim18b}
\bibinfo{author}{\bibfnamefont{X.}~\bibnamefont{Liu}},
  \bibinfo{author}{\bibfnamefont{Z.}~\bibnamefont{Hao}},
  \bibinfo{author}{\bibfnamefont{K.}~\bibnamefont{Watanabe}},
  \bibinfo{author}{\bibfnamefont{T.}~\bibnamefont{Taniguchi}},
  \bibinfo{author}{\bibfnamefont{B.}~\bibnamefont{Halperin}}, \bibnamefont{and}
  \bibinfo{author}{\bibfnamefont{P.}~\bibnamefont{Kim}},
  \bibinfo{journal}{arXiv e-prints} \bibinfo{eid}{arXiv:1810.08681}
  (\bibinfo{year}{2018}), \eprint{arXiv:1810.08681}.

\bibitem[{\citenamefont{Li et~al.}(2019)\citenamefont{Li, Shi, Y.~Zeng,
  Taniguchi, Hone, and Dean}}]{Dean18}
\bibinfo{author}{\bibfnamefont{J.}~\bibnamefont{Li}},
  \bibinfo{author}{\bibfnamefont{Q.}~\bibnamefont{Shi}},
  \bibinfo{author}{\bibfnamefont{K.~W.} \bibnamefont{Y.~Zeng}},
  \bibinfo{author}{\bibfnamefont{T.}~\bibnamefont{Taniguchi}},
  \bibinfo{author}{\bibfnamefont{J.}~\bibnamefont{Hone}}, \bibnamefont{and}
  \bibinfo{author}{\bibfnamefont{C.}~\bibnamefont{Dean}},
  \bibinfo{journal}{arXiv e-prints} \bibinfo{eid}{arXiv:1901.03480}
  (\bibinfo{year}{2019}), \eprint{arXiv:1901.03480}.

\bibitem[{\citenamefont{Kane and Fisher}(1997)}]{Kane97}
\bibinfo{author}{\bibfnamefont{C.~L.} \bibnamefont{Kane}} \bibnamefont{and}
  \bibinfo{author}{\bibfnamefont{M.~P.~A.} \bibnamefont{Fisher}},
  \bibinfo{journal}{Phys. Rev. B} \textbf{\bibinfo{volume}{55}},
  \bibinfo{pages}{15832} (\bibinfo{year}{1997}),
  \urlprefix\url{http://link.aps.org/doi/10.1103/PhysRevB.55.15832}.

\bibitem[{\citenamefont{Chang et~al.}(1996)\citenamefont{Chang, Pfeiffer, and
  West}}]{Chang96}
\bibinfo{author}{\bibfnamefont{A.~M.} \bibnamefont{Chang}},
  \bibinfo{author}{\bibfnamefont{L.~N.} \bibnamefont{Pfeiffer}},
  \bibnamefont{and} \bibinfo{author}{\bibfnamefont{K.~W.} \bibnamefont{West}},
  \bibinfo{journal}{Phys. Rev. Lett.} \textbf{\bibinfo{volume}{77}},
  \bibinfo{pages}{2538} (\bibinfo{year}{1996}),
  \urlprefix\url{https://link.aps.org/doi/10.1103/PhysRevLett.77.2538}.

\bibitem[{\citenamefont{Grayson et~al.}(1998)\citenamefont{Grayson, Tsui,
  Pfeiffer, West, and Chang}}]{Grayson98}
\bibinfo{author}{\bibfnamefont{M.}~\bibnamefont{Grayson}},
  \bibinfo{author}{\bibfnamefont{D.~C.} \bibnamefont{Tsui}},
  \bibinfo{author}{\bibfnamefont{L.~N.} \bibnamefont{Pfeiffer}},
  \bibinfo{author}{\bibfnamefont{K.~W.} \bibnamefont{West}}, \bibnamefont{and}
  \bibinfo{author}{\bibfnamefont{A.~M.} \bibnamefont{Chang}},
  \bibinfo{journal}{Phys. Rev. Lett.} \textbf{\bibinfo{volume}{80}},
  \bibinfo{pages}{1062} (\bibinfo{year}{1998}),
  \urlprefix\url{https://link.aps.org/doi/10.1103/PhysRevLett.80.1062}.

\bibitem[{\citenamefont{Chang}(2003)}]{Chang03}
\bibinfo{author}{\bibfnamefont{A.~M.} \bibnamefont{Chang}},
  \bibinfo{journal}{Rev. Mod. Phys.} \textbf{\bibinfo{volume}{75}},
  \bibinfo{pages}{1449} (\bibinfo{year}{2003}),
  \urlprefix\url{http://link.aps.org/doi/10.1103/RevModPhys.75.1449}.

\bibitem[{\citenamefont{Read}(2009)}]{Read09}
\bibinfo{author}{\bibfnamefont{N.}~\bibnamefont{Read}}, \bibinfo{journal}{Phys.
  Rev. B} \textbf{\bibinfo{volume}{79}}, \bibinfo{pages}{045308}
  (\bibinfo{year}{2009}),
  \urlprefix\url{http://link.aps.org/doi/10.1103/PhysRevB.79.045308}.

\bibitem[{\citenamefont{Morf}(1986)}]{Morf86}
\bibinfo{author}{\bibfnamefont{R.}~\bibnamefont{Morf}},
  \bibinfo{author}{\bibfnamefont{N.} \bibnamefont{d'Ambrumenil}},
  \bibinfo{author}{\bibfnamefont{B.~I.} \bibnamefont{Halperin}}, \bibinfo{journal}{Phys.
  Rev. B} \textbf{\bibinfo{volume}{34}}, \bibinfo{pages}{3037}
  (\bibinfo{year}{1986}),
  \urlprefix\url{http://link.aps.org/doi/10.1103/PhysRevB.34.3037}.

\bibitem[{\citenamefont{Wen}(1992)}]{Wen92}
\bibinfo{author}{\bibfnamefont{X.~G.}~\bibnamefont{Wen}},
  \bibinfo{author}{\bibfnamefont{A.} \bibnamefont{Zee}}, \bibinfo{journal}{Phys.
  Rev. Lett.} \textbf{\bibinfo{volume}{69}}, \bibinfo{pages}{953}
  (\bibinfo{year}{1992}),
  \urlprefix\url{http://link.aps.org/doi/10.1103/PhysRevLett.69.953}.

\bibitem[{\citenamefont{Halperin}(1984)}]{Halperin84}
\bibinfo{author}{\bibfnamefont{B.~I.}~\bibnamefont{Halperin}}, \bibinfo{journal}{Phys.
  Rev. Lett} \textbf{\bibinfo{volume}{52}}, \bibinfo{pages}{1583}
  (\bibinfo{year}{1984}),
  \urlprefix\url{http://link.aps.org/doi/10.1103/PhysRevLett.52.1583}.

\bibitem[{\citenamefont{Belkhir}(1993)}]{Belkhir93}
\bibinfo{author}{\bibfnamefont{L.}~\bibnamefont{Belkhir}},
  \bibinfo{author}{\bibfnamefont{X.~G.} \bibnamefont{Wu}},
  \bibinfo{author}{\bibfnamefont{J.~K.} \bibnamefont{Jain}}, \bibinfo{journal}{Phys.
  Rev. B} \textbf{\bibinfo{volume}{48}}, \bibinfo{pages}{15245}
  (\bibinfo{year}{1993}),
  \urlprefix\url{http://link.aps.org/doi/10.1103/PhysRevB.48.15245}.

\end{thebibliography}
%\bibliographystyle{apsrev}
%\end{document}

\end{document}